\documentclass[fleqn,10pt]{wlscirep}
\usepackage[utf8]{inputenc}
\usepackage[T1]{fontenc}
\usepackage{xcolor}
\usepackage[table]{xcolor}
\usepackage{algpseudocode}
\usepackage{graphicx}
\usepackage{textcomp}
\usepackage{xcolor}

\newcommand{\black}[1]{\textcolor{black}{#1}}
\definecolor{darkgreen}{rgb}{0.0, .75, 0.0} 
\usepackage{pgfplots}
\usepackage{color,soul}
\usepackage{booktabs}
\usepackage{vcell}
\usepackage{lipsum}
\usepackage{tabularx}
\usepackage{adjustbox}
\usepackage{romannum}
\usepackage{longtable}
\usepackage{multirow}
\usepackage{array}
\usepackage[font=footnotesize]{caption}
\usepackage[font=footnotesize]{subcaption}
\usepackage{pdflscape}
\usepackage{tabu}
\usepackage{supertabular}
\usepackage{makecell}
\usepackage{tabu}
\usepackage[ruled, linesnumbered, longend]{algorithm2e}
\usepackage{array}
\usepackage[font=footnotesize]{subcaption}
\usepackage{textcomp}
\usepackage{stfloats}
\usepackage{url}
\usepackage{verbatim}
\usepackage{cite}
\usepackage{adjustbox}
\usepackage{booktabs}
\usepackage{makecell}
\usepackage{array}

\newcolumntype{C}[1]{>{\centering\arraybackslash}m{#1}}  
\title{Low-Latency Neural Inference on an Edge Device for Real-Time Handwriting Recognition from EEG Signals}

\author[1,*]{Ovishake Sen}
\author[2]{Raghav Soni}
\author[1]{Darpan Virmani}
\author[3]{Akshar Parekh}
\author[1]{Patrick Lehman}
\author[2]{Sarthak Jena}
\author[1]{Adithi Katikhaneni}
\author[1]{Adam Khalifa}
\author[1]{Baibhab Chatterjee}
\affil[1]{Electrical and Computer Engineering, University of Florida, Gainesville, FL 32611, USA}
\affil[2]{Computer and Information Science and Engineering, University of Florida, Gainesville, FL 32611, USA}
\affil[3]{Behavioral and Cognitive Neuroscience, University of Florida, Gainesville, FL 32611, USA}

\affil[*]{corresponding author: ovishake.sen@ufl.edu}

\begin{abstract}
Brain–computer interfaces (BCIs) hold significant promise for restoring communication in individuals with severe motor or speech impairments. Imagined handwriting, as a form of motor imagery, offers an intuitive paradigm for character-level neural decoding. While invasive techniques such as electrocorticography (ECoG) offer high decoding accuracy, their surgical requirements pose clinical risks and hinder scalability. Non-invasive alternatives like electroencephalography (EEG) are safer and more accessible but suffer from low signal-to-noise ratio (SNR) and spatial resolution, limiting their effectiveness in high-resolution decoding. Here, we investigate how advanced machine learning, combined with informative feature extraction, can overcome these limitations—enabling EEG-based decoding performance that approaches invasive methods, while supporting real-time inference on edge devices. We present the first real-time, low-latency, high-accuracy system for decoding imagined handwriting from non-invasive EEG signals on a portable edge device. \black{EEG data were collected from 15 participants using a 32-channel headcap and preprocessed with bandpass filtering and artifact subspace reconstruction (ASR).} \black{We extracted 85 time- domain, frequency- domain, and graphical features, then applied Pearson correlation coefficient-based feature selection to reduce latency while preserving accuracy}.
\black{We developed a hybrid architecture, EEdGeNet, which integrates a Temporal Convolutional Network (TCN) with a multilayer perceptron (MLP), trained on the extracted features and deployed on the NVIDIA Jetson TX2 for real-time inference.} \black{The system achieved $89.83\% \pm 0.19\%$ accuracy with 914.18 ms per-character inference latency. By selecting only ten key features, the model incurred a minimal accuracy loss of $<1\%$, while achieving a $4.51\times$ reduction in inference latency (202.62 ms) compared to the full 85-feature set.} These findings show that non-invasive EEG, combined with efficient feature and model design, can enable accurate, real-time neural decoding on low-power edge devices—paving the way for practical, portable BCIs.
\end{abstract}

\begin{document}

\flushbottom
\maketitle
%
%
\thispagestyle{empty}


\section*{Introduction}
Imagined handwriting offers an intuitive and expressive mechanism for translating neural intent into character-level text. This paradigm leverages well-established motor representations in the brain and enables high-throughput decoding, making it particularly promising for individuals with severe motor or speech impairments \cite{zhang2024brain}.

While invasive modalities such as ECoG and intracortical microelectrode arrays provide high signal fidelity and spatial resolution, their reliance on surgical implantation imposes clinical risks and limits long-term adoption. Non-invasive techniques such as EEG are safer, more accessible, and better suited for scalable applications. EEG-based motor imagery systems have shown promise in assistive domains, including speller interfaces, virtual cursor control, and thought-to-text conversion \cite{liu2025recent}. However, EEG signals are inherently noisy, non-stationary, and spatially coarse—posing challenges for reliable decoding in fine-grained tasks like handwriting \cite{sun2021hybrid}.

Recent advances in machine learning and signal processing have begun to address these limitations. Techniques such as artifact subspace reconstruction, temporal-spectral filtering, and targeted feature extraction can enhance signal quality and reduce dimensionality while preserving meaningful neural dynamics. When combined with compact neural architectures, these strategies enable decoding performance that increasingly approaches invasive systems, even from non-invasive data \cite{ahmadlou2012fractality,sun2021hybrid}.

For practical deployment, BCIs must operate in real time on portable, low-power platforms. On-device inference reduces latency, improves responsiveness, and eliminates dependence on external processing, which is especially critical for users with mobility constraints \cite{janapa2022mlperf, hadidi2020toward}. However, edge deployment of EEG-based decoders remains challenging due to the computational cost of deep models and the high dimensionality of raw signals \cite{safari2024classification, petrosyan2022speech}. Efficient preprocessing and feature extraction are thus essential—not only to suppress artifacts and noise, but also to retain physiologically meaningful components that reduce computational overhead and enable robust, low-latency inference on resource-constrained hardware.

\begin{figure}[t]
    \centering
    \includegraphics[width=\linewidth]{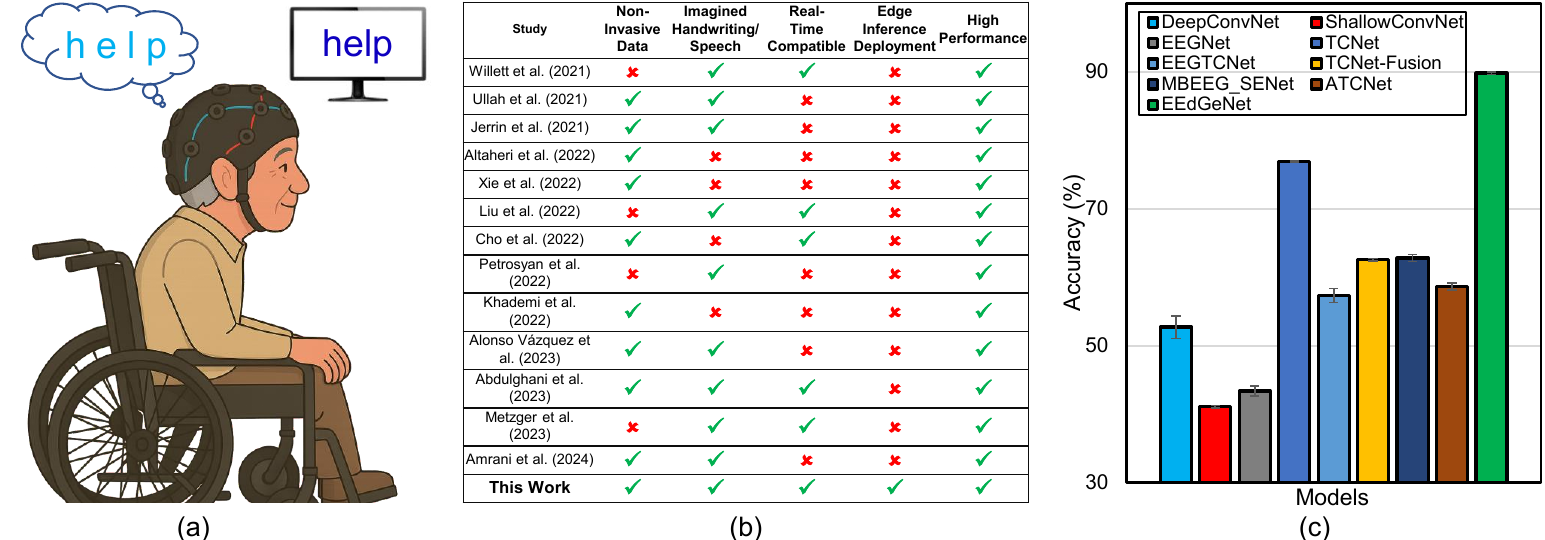}
    \caption{Proposed framework for real-time imagined handwriting decoding from non-invasive EEG signals.
(a) Illustration of a motor-impaired person wearing a multi-channel EEG headcap for real-time neural decoding. (b) Comparative analysis of prior studies based on: non-invasive signal acquisition, imagined handwriting/speech decoding, real-time inference compatibility, edge-level deployment feasibility, and classification performance. The proposed approach is the only method that meets all criteria concurrently. \black{(c) Accuracy comparison with state-of-the-art EEG models on 85 features extracted from imagined handwriting data, showing the superior performance of our proposed model (EEdGeNet) for edge-based inference.}}
    \label{fig:motivation}
\end{figure}

Recent works on character and speech recognition using invasive signals relying on intracortical or implanted electrode systems have utilized Utah arrays, stereoelectroencephalography (SEEG), and ECoG to achieve high-resolution recording of neural signals \cite{liu2022intracranial, petrosyan2022speech, metzger2023high}. These systems have been used for the classification of spoken or attempted speech and handwriting. Willett et al. used Utah arrays implanted in the motor cortex to decode attempted handwriting, achieving 90 characters per minute with 94.1\% accuracy in real time \cite{willett2021high}. Other works have utilized deep learning models, including recurrent neural networks and encoder-decoder architectures to achieve high classification accuracies for signal decoding. Although these methods demonstrate high real-time performance, they require surgical implantation and clinical oversight \cite{liu2022intracranial, petrosyan2022speech, metzger2023high,willett2021high}. In contrast, our approach leverages non-invasive EEG to enable real-time, low-latency decoding of imagined handwriting on edge devices, offering a safer and more accessible alternative.

For non-invasive EEG-based decoding, studies have employed diverse preprocessing and artifact removal pipelines—including usage of bandpass and notch filtering, line noise removal, baseline correction, Independent Component Analysis (ICA), Artifact Subspace Reconstruction (ASR), Multiple Artifact Rejection Algorithm (MARA), and wavelet or entropy-based transformations—to mitigate artifacts and enhance signal quality for downstream classification tasks. Several works also incorporate referencing schemes, normalization (such as Z-score), and dimensionality reduction techniques and follow preprocessing with feature extraction \cite{mishra2024signeeg, safari2024classification, sun2021hybrid, zhao2021classification}. These preprocessing pipelines have been applied for diverse EEG classification tasks such as Alzheimer's stage classification \cite{kim2024electroencephalography}, speech and vocabulary generation \cite{amrani2024deep} and ischemic stroke classification \cite{islam2022explainable}. This shows that usage of such preprocessing pipelines is critical for preparing clean and informative data for EEG classification.

In imagined character and speech recognition, most EEG-based studies have predominantly relied on offline deep learning models, achieving high classification accuracy without addressing real-time inference. Ullah et al. (2021) focused on imagined character recognition using a deep Convolutional Neural Network (CNN) trained on Morlet wavelet-transformed EEG signals \cite{ullah2021imagined}. Their model was validated on the EEGMMIDB and BCI Competition IV-2a datasets, as well as a custom visual-imagery dataset comprising 10 subjects and 26 English characters. They reported classification accuracies of 99.45\% on public datasets and 95.2\% on their proprietary dataset. However, the model was not evaluated for real-time inference capability. Several studies have explored EEG decoding using various other deep learning models and traditional machine learning techniques. Models and architectures such as SVMs, LSTMs, CNN-based EEGNet, pretrained models on ResNet50, BART-based LLMs to capture spatiotemporal patterns and generalize across complex EEG data have been used for speech and character classification. Various feature extraction pipelines have also been incorporated with these models that extract specific features ranging from spectral power and coherence to wavelet-scattering and 3D topographic representations achieving classification accuracies up to 95.5\% in specific tasks \cite{alonso2023eeg, amrani2024deep, abdulghani2023imagined, panachakel2021decoding}. Despite having high offline performance, most of these frameworks are have been designed and evaluated in non-real-time settings, limiting their current applicability for real-time deployment.

For motor imagery EEG classification, studies have primarily utilized offline deep learning frameworks—such as CNNs, LSTMs, and transformers—with physics-informed feature extraction, spatial-temporal attention-based learning or hybrid architectures, achieving high accuracy but limited real-time applicability. These models have been employed to classify binary, three-class, and four-class motor imagery tasks and have been validated using BCI Competition datasets, with several achieving offline accuracies up to 92\% \cite{altaheri2022physics, xie2022transformer, khademi2022transfer}. While most frameworks have been developed and evaluated in offline mode, NeuroGrasp, has demonstrated partial real-time feasibility using CNN-BLSTM and Siamese networks \cite{cho2021neurograsp}. Overall, these architectures are designed to maximize accuracy by learning intricate signal representations, though real-time applicability remains limited in most cases.

\textcolor{black}{Traditional BCI paradigms like Steady-State Visual Evoked Potential (SSVEP) based BCIs achieve high information transfer rates \cite{nakanishi2017enhancing}, but they depend on sustained visual attention to flickering stimuli, which can induce fatigue and are
unsuitable for patients with visual or oculomotor impairments. Event-Related Potential (ERP)/P300 spellers offer reliable performance \cite{farwell1988talking}, yet they also rely on repeated stimulus presentation, leading to slower communication speeds and reduced practicality for continuous use. Speech-attempt paradigms, including overt or attempted articulation, provide natural mapping to language but typically require invasive recordings (ECoG or intracortical arrays) to reach state-of-the-art performance \cite{metzger2023high,willett2021high}.  
In contrast, imagined handwriting directly encodes rich, character-level linguistic information without reliance on external stimuli or residual motor ability.}

Figure \ref{fig:motivation}a illustrates the proposed framework for real-time neural inference from non-invasive EEG signals to decode imagined handwriting. A motor- and communication-impaired individual is equipped with a multi-channel EEG headcap, and the system is designed to translate character-level motor intentions into text in real-time through low-latency inference pipelines deployed on edge hardware. This approach demonstrates the feasibility of portable neural decoding, offering high-speed processing and accurate predictions directly on resource-constrained devices. As shown in Figure \ref{fig:motivation}b, unlike conventional methods that depend on cloud-based infrastructure, high-performance computing platforms, or invasive neural recordings, our system enables on-device computation with non-invasive EEG data, significantly reducing latency and enhancing deployment practicality. \black{Comparative validation against existing EEG-based models, as depicted in Figure \ref{fig:motivation}c, demonstrates that the proposed EEdGeNet architecture achieves superior performance in decoding imagined handwritten characters from feature-extracted EEG signals.}

In this paper, we present a novel methodology for real-time inference from non-invasive EEG signals to decode imagined handwriting. The key contributions of this study are as follows:
\begin{itemize}
\item \black{We demonstrate, for the first time, real-time inference of imagined handwriting from non-invasive EEG signals, achieving a character classification accuracy of $89.83\% \pm 0.19\%$ using 85 extracted features from raw EEG data.}
\item We implement, for the first time, a real-time EEG decoding pipeline fully deployed on a portable hardware platform (NVIDIA Jetson TX2), with an average inference latency of 914.18 ms per-character.
\item \black{We select ten key features using Pearson correlation, achieving $88.84\% \pm 0.09\%$ accuracy while reducing inference latency by $4.51\times$ compared to the full feature set (202.62 ms), enabling real-time BCI with minimal computational overhead.}
\item \black{We develop EEdGeNet, a hybrid TCN–MLP architecture that can be efficiently deployed on the NVIDIA Jetson TX2 and achieves superior performance in imagined handwriting recognition compared to state-of-the-art EEG classifiers.}
\item Our optimized feature extraction pipeline, combined with a lightweight and tuned neural model that avoids the high computational overhead typical of existing motor imagery classifiers, enables real-time inference and deployment on resource-constrained edge devices.
\end{itemize}
\black{In essence, this work presents the first real-time, high-accuracy decoding of imagined handwriting from non-invasive EEG signals on a portable edge device, demonstrating robust generalizability across multiple participants and low-latency performance suitable for assistive communication. The proposed method achieves $89.83\%\pm 0.19\%$ accuracy across 15 participants using 85 extracted features, with an inference latency of 914.18 ms per-character on the NVIDIA Jetson TX2. Selecting only ten key features results in $<1\%$ accuracy loss while reducing latency by $4.51\times$, demonstrating the feasibility of low-power, real-time BCI systems for assistive communication.}
\section*{Materials and Methods}

\subsection*{The proposed real-time EEG-based handwriting recognition pipeline}
\begin{figure}[htbp]
    \centering
    \includegraphics[width=\linewidth]{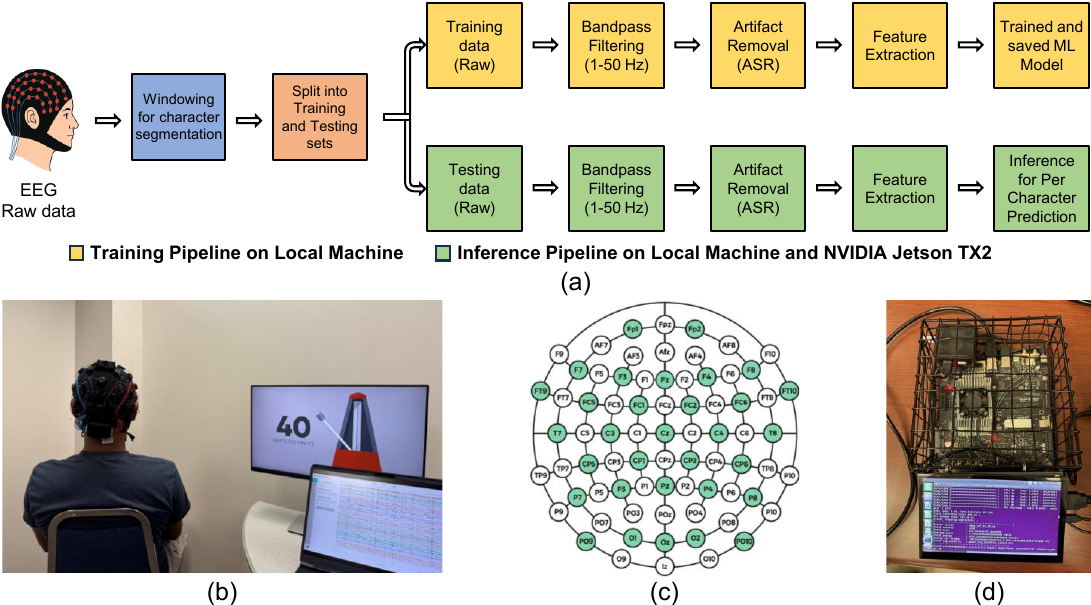}
    \caption{Proposed methodology for low-latency neural inference on an edge device for real-time handwriting recognition from EEG signals.
    (a) The entire pipeline for training (yellow) and inference (green), including preprocessing, feature extraction, and per-character prediction. (b) Data collection setup showing a participant wearing the Emotiv Flex 2.0 EEG headcap during data acquisition. (c) Electrode layout based on the international 10–20 system, with utilized channels highlighted in green. (d) Real-time inference on the NVIDIA Jetson TX2 with a batch size of 1.}
    \label{fig:methodology}
\end{figure}
\black{The methodology for decoding imagined handwritten characters from EEG signals began with data acquisition from 15 participants. Raw EEG recordings were segmented into per-character windows and preprocessed to suppress noise and artifacts. From each segment, 85 features were then extracted, covering the time and frequency domains, along with graphical descriptors. To minimize computational overhead during inference, a feature selection step is performed, retaining only the most discriminative features and spatially relevant EEG channels associated with handwriting intent. The feature set is then used to train a suite of ML classifiers, with model performance assessed on a held-out test set to evaluate generalizability.} 
The complete framework for low-latency, real-time EEG inference is illustrated in Figure \ref{fig:methodology}. Figure \ref{fig:methodology}a presents the end-to-end pipeline, including bandpass filtering, ASR-based artifact removal, feature extraction, and model execution for both training (on a local machine) and inference (on local machine as well as on edge hardware). Figure \ref{fig:methodology}b shows the EEG data acquisition setup with the Emotiv headset. Figure \ref{fig:methodology}c illustrates the electrode placement following the international 10–20 system, with the channels used in our study highlighted in green. Figure \ref{fig:methodology}d demonstrates real-time deployment on the NVIDIA Jetson TX2 for per-character prediction. The following sections present a detailed overview of the key components comprising the proposed pipeline.

\subsection*{Participants and EEG recording}
\black{EEG signals were collected from 15 participants using the Emotiv Epoch FLEX 2.0 saline-based head cap, comprising 32 scalp electrodes with a sampling frequency of 256 Hz.} All experimental protocols were approved by the University of Florida Institutional Review Board (IRB202301691) and were conducted in accordance with the relevant guidelines and regulations. Informed consent was obtained from all participants. Each participant was instructed to imagine writing all 26 lowercase English alphabet characters (a–z), in addition to one “do nothing phase” intended to capture resting-state neural activity without any intentional motor or cognitive imagery.
All recording sessions were conducted in a quiet room to minimize environmental artifacts and auditory or visual distractions. Prior to data acquisition, participants were instructed to relax and avoid interaction with any electronic devices to reduce extraneous neural activity. The recording session commenced with a structured baseline protocol consisting of a 5-second preparation window, followed by 15 seconds of eyes-open relaxation, a 5-second transition period, and then 15 seconds of eyes-closed relaxation. A subsequent 20-second resting period was given before the onset of the data recording phase.

Imagined handwriting EEG data collection was synchronized to a 40 BPM metronome. Starting from the fourth beat (1 minute and 6 seconds into the session), participants were instructed to begin imagining the act of writing the target character in synchrony with each subsequent metronome beat. Data was recorded continuously until the 144th beat (4 minutes and 36 seconds). Participants were then given a short break before proceeding with the next character trial.
The complete set of 26 alphabet characters and the do nothing phase was recorded over multiple sessions (typically 4–5), adjusted based on each participant’s cognitive load and relaxation level. Each session included data collection for 6 to 8 characters. Following each session, the raw EEG signals were exported as CSV files using EmotivPRO software for subsequent preprocessing and analysis.

\subsection*{Data preprocessing}
Following data acquisition, the raw EEG signals collected from each participant were subjected to a standardized preprocessing pipeline designed to attenuate noise and artifacts while preserving the natural neural activity. To ensure compatibility with real-time inference, preprocessing techniques were applied separately to each character window after the train/test split. This design reflects the real-time constraint of obtaining per-character predictions, necessitating that preprocessing be performed independently on each window.  


\black{Accordingly, while independent component analysis (ICA), principal component analysis (PCA), discrete wavelet transform (DWT), empirical wavelet transform (EWT), and variational mode decomposition (VMD) were explored, these methods proved insufficient for fully denoising the signal—likely due to the shorter window lengths inherent to character-level EEG segments~\cite{huotari2019sampling,beckmann2005investigations}.} We therefore employed ASR, a method demonstrated to be effective for cleaning short-duration EEG recordings~\cite{chang2019evaluation,blum2019riemannian}. ASR balances robustness with computational efficiency, making it particularly well suited for real-time deployment on edge devices. The preprocessing steps are outlined as follows.

\begin{figure}[!b]

\begin{subfigure}[b]{0.495\linewidth}
\centering
\captionsetup{justification=centering}
    \includegraphics[width=\linewidth]{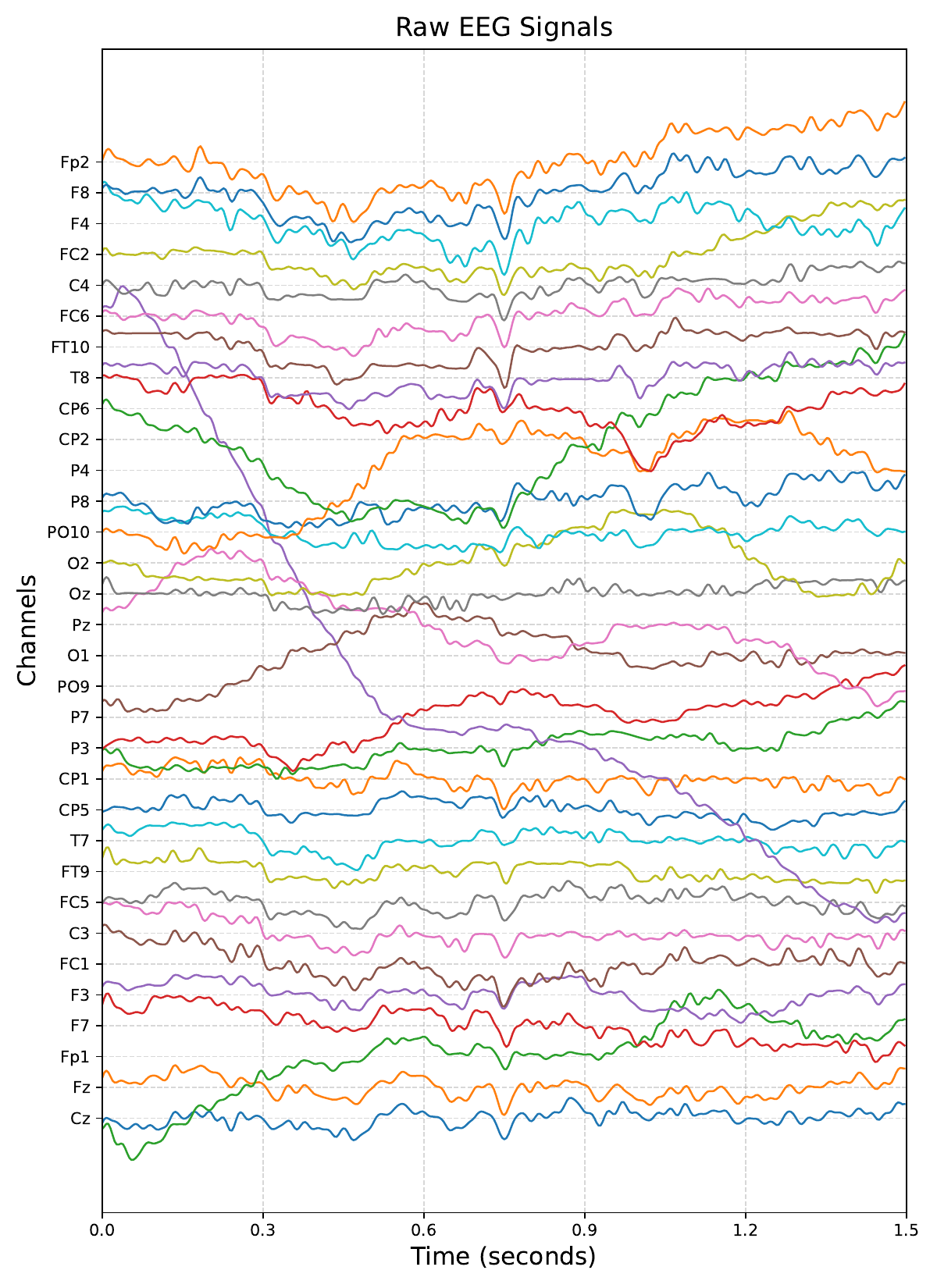}
    \caption{}
\label{fig:rawdata}
\end{subfigure}
\begin{subfigure}[b]{0.495\linewidth}
\centering
\captionsetup{justification=centering}
    \includegraphics[width=\linewidth]{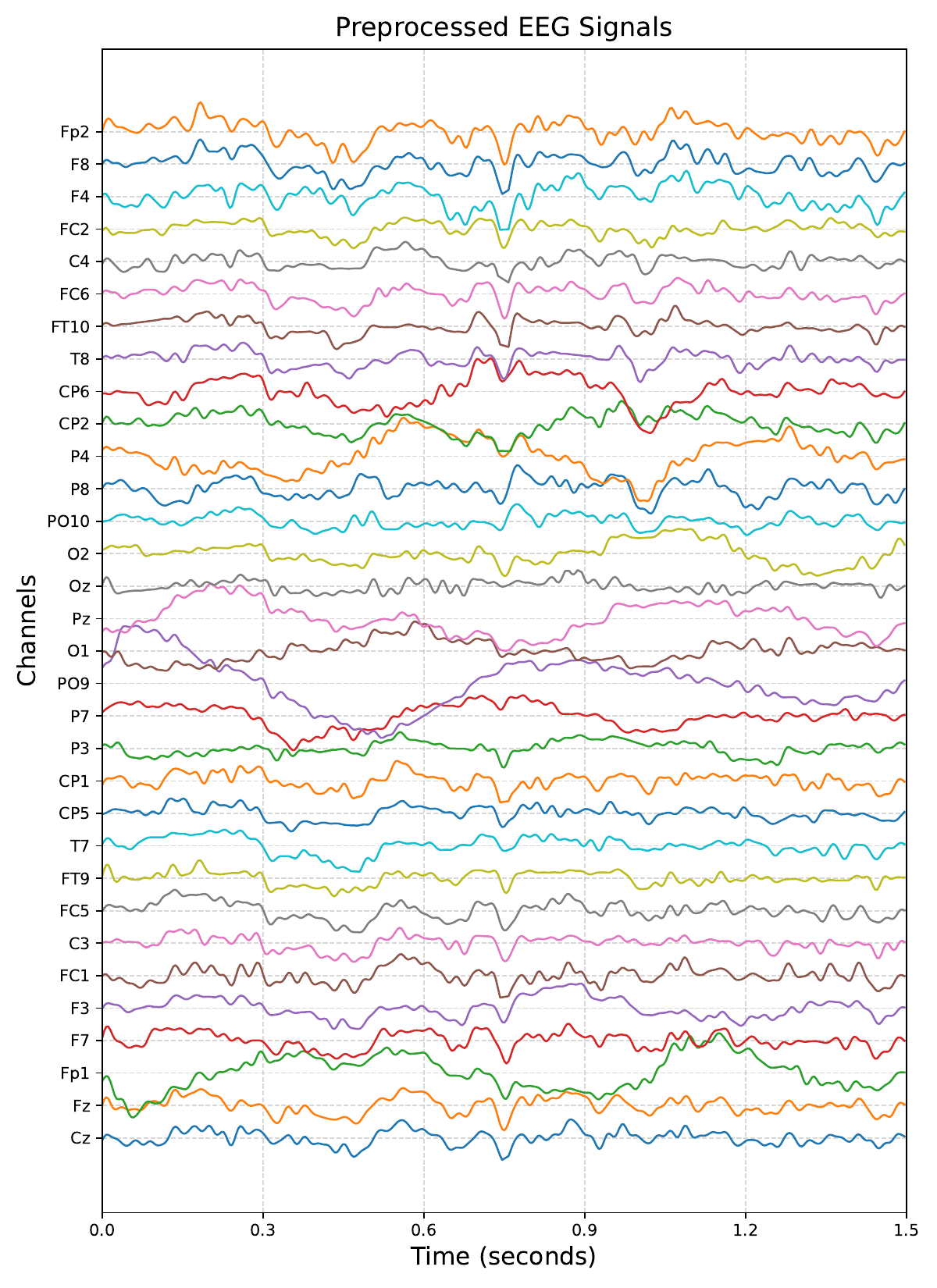}
    \caption{}
    \label{fig:preprocessed}
\end{subfigure}
    \caption{(a) Raw EEG signals. (b) Preprocessed EEG signals. The 32-channel EEG data from a participant’s sample for the character \textit{'v'} is shown before and after preprocessing. Bandpass filtering (1–50 Hz) and Artifact Subspace Reconstruction (ASR) were applied to effectively remove noise and artifacts, improving overall signal quality.}
\label{fig:preprocessing}

\end{figure}
\begin{enumerate}
    \item \textbf{Bandpass filtering} To remove low-frequency drifts and high-frequency noise, the raw EEG signals were passed through a bandpass filter with a passband of 1–50\,Hz\cite{Yeh2025Corticomuscular, Mahadevan2024EEGfMRI, Choi2018EarEEG}. This range captures the relevant EEG frequency bands while excluding low-frequency noise arising from sources such as head movements, electrode wire displacement, scalp perspiration, or slow drifts in the EEG signal over extended periods \cite{safari2024classification}. A fourth-order Butterworth filter was designed using the \texttt{scipy.signal.butter} function and applied uniformly across all channels for every EEG sample.
    
    \item \textbf{ASR} After bandpass filtering, we employed a simplified variant of Artifact Subspace Reconstruction (ASR) to suppress transient, high-amplitude disturbances. This method was applied independently to each EEG channel. For each channel, the mean and standard deviation were computed, and samples exceeding a z-score threshold of 3.0 were identified as artifacts and replaced with NaN values. Missing values were subsequently reconstructed using linear interpolation, followed by backward and forward filling to ensure signal continuity at temporal boundaries. This procedure preserved the intrinsic temporal structure of the EEG signal while mitigating large, non-stationary artifacts \cite{mullen2015real}.
\end{enumerate}

Figure \ref{fig:rawdata} illustrates a participant's 1.5-second EEG segment recorded during the writing of the character \textit{'v'}, while Figure \ref{fig:preprocessed} shows the corresponding signal following preprocessing, demonstrating the effective removal of noise and artifacts.

\subsection*{Feature extraction}
\textcolor{black}{Following preprocessing, a comprehensive set of features was extracted from each EEG window to capture both linear and non-linear signal characteristics.}
Specifically, we computed 12 time-domain features, 8 frequency-domain features, and \textcolor{black}{65 graphical features derived from Poincaré plots of discrete wavelet transform sub-bands, as described in \cite{akbari2023recognizing}}.
The time-domain features include: mean, variance, skewness, kurtosis, Root Mean Square (RMS), Slope Sign Changes (SSC), Hjorth mobility, Hjorth complexity, Hurst exponent, Shannon entropy, sample entropy, and permutation entropy.
The frequency-domain features consist of: spectral band powers in the delta, theta, alpha, beta, and gamma bands, peak frequency, mean dominant frequency, and spectral entropy. 
\textcolor{black}{The graphical features comprised thirteen descriptors computed from each Poincaré plot: ellipse area, eccentricity, first and second eigenvalues, mean deviation from the identity line, mean distance from the origin, sequential standard deviation of vector lengths (SSVL), covariance, mean x- and y-coordinates, standard deviations along both axes, and the central tendency measure (CTM). These features were extracted independently from the approximation (A) and detail (D1–D4) sub-bands of the DWT, resulting in a total of 65 graphical features per EEG window.}


\subsubsection*{Time domain features}
Mean, RMS, SSC, and Variance were calculated using the NumPy library. Skewness, kurtosis, and Shannon entropy were computed using the SciPy library. Hjorth mobility and Hjorth complexity were calculated as described by \cite{hjorth1970eeg}. Permutation entropy was calculated as described by \cite{bandt2002permutation}. 

Hurst Exponent \cite{pernagallo2025random} was computed using the equation \eqref{hurst}:
\begin{equation}
Hurst\_Exponent = \frac{ \log(R/S) - \log c }{ \log N } \label{hurst}
\end{equation}

where \( R \) is the range of the cumulative sum of deviations of the signal values from the mean, \( S \) is the standard deviation, and \( N \) is the number of samples of the signal. In EEG analysis, the constant c only shifts the intercept and does not affect the Hurst exponent, so it is set to 1 to simplify computation and avoid unstable estimation in noisy signals.

Sample Entropy was computed using the equation \eqref{sample_entropy}:

\begin{equation}
\mathrm{SampEn}(m, r, N) = -\ln \left( \frac{\Phi^{m+1}(r)}{\Phi^m(r)} \right) \label{sample_entropy}
\end{equation}

Where \( m \) is the length of the subsequence, \( r \) is the tolerance, \( N \) is the number of data points in the signal, and \( \varphi_m(r) \) is the probability that two subsequences of length \( m \) are similar, where similarity is determined by the tolerance \( r \).

where Probability of a match between two subsequences within the signal is defined as:

\[
\Phi^m(r) = \frac{1}{N - m + 1} \sum_{i=1}^{N - m + 1} \frac{B^m_i(r)}{N - m}
\]

$B^m_i(r)$ is the count of pairs of subsequences for which the Chebyshev distance between that pair is less than the tolerance parameter \( r \) defined as \( r = 0.2 \times \sigma \), where \( \sigma \) is the standard deviation of the signal. Chebyshev distance between two subsequences within the signal is defined as:

\[
d[\mathbf{u}(i), \mathbf{u}(j)] = \max_{k=0,\dots,m-1} |x_{i+k} - x_{j+k}|
\]
where $\mathbf{u}(i) = \{x_i, x_{i+1}, \dots, x_{i+m-1}\}$ is a subsequence of length $m$ from a signal of length $N$

We use \( N - m + 1 \) instead of \( N - m \) because it accurately reflects the total number of embedding vectors of length \( m \) in a time series of length \( N \). This avoids excluding the final valid segment and prevents systematic underestimation in entropy computation.

\textcolor{black}{\subsection*{Graphical features from Poincaré-DWT analysis}
Graphical features were extracted from Poincaré plots constructed using the DWT coefficients of each EEG channel within a 1.5-second window. The EEG signal was decomposed into five sub-bands—approximation (A) and detail levels (D1–D4)—using the \texttt{wavedec} function from the PyWavelets library. For each sub-band, time-lagged DWT coefficients were embedded in phase space to generate Poincaré plots, capturing the dynamical evolution of the signal over time.
Thirteen graphical descriptors were computed from each plot using NumPy, characterizing the shape, spread, and structural complexity of the trajectory: ellipse area, eccentricity, first and second eigenvalues (representing variances along the principal axes), mean deviation from the identity line, mean distance from the origin, sequential standard deviation of vector lengths (SSVL), covariance, mean of x- and y-coordinates, standard deviations along both axes, and central tendency measure (CTM). Applied across all five DWT sub-bands, this resulted in 65 graphical features per EEG window.
These non-linear descriptors capture morphological and recurrence-based patterns in neural activity and have been shown to enhance classification performance in a range of cognitive and neurological decoding tasks \cite{akbari2023recognizing,sadiq2022alcoholic,akbari2021depression}.}

\noindent
\subsection*{Feature selection}
We used the Pearson correlation coefficient to select the most important features, calculated using the equation \eqref{eq:pearson}.

\begin{equation}
r = \frac{\sum_{i=1}^{n} (x_i - \bar{x})(y_i - \bar{y})}
         {\sqrt{\sum_{i=1}^{n} (x_i - \bar{x})^2} \cdot \sqrt{\sum_{i=1}^{n} (y_i - \bar{y})^2}}
         \label{eq:pearson}
\end{equation}

Here, $r$ is the Pearson correlation coefficient. $x_i$, $y_i$ are values of all features \textcolor{black}{(85)} and $n$ is the number of channels (32).
$r$ measures the linear relationship between all the feature pairs. Where
$r = 1$: perfect positive correlation, $r = -1$: perfect negative correlation, $r = 0$: no linear correlation.
To identify the most relevant features, we first included negatively correlated feature pairs with a correlation less than \textcolor{black}{-0.4}. Then, if any selected feature pairs exhibited a correlation greater than \textcolor{black}{0.7}, we discarded one feature from each such pair to avoid redundancy. \textcolor{black}{Finally, we selected the 10 most important features that provided approximately the same accuracy as using all 85 features, while improving inference time/character by $4.51\times$ and inference energy/character by $5.41\times$}.

\section*{Evaluation Methodology}
\subsection*{{Proposed \black{EEdGeNet} Model architecture}}

The proposed \black{EEdGeNet} model architecture comprises two principal blocks: a Temporal Convolutional Block and a Dense Transformation Block. The first block employs a Temporal Convolutional Network (TCN) to effectively capture both spatial and temporal dependencies inherent in the features extracted from raw EEG signals. The second block flattens the temporal representations and transforms them into a compact feature vector, leveraging a deep Multilayer Perceptron (MLP) to learn complex non-linear mappings. The outputs from all sliding windows are concatenated and fed into a softmax classifier to perform multi-class classification. Further architectural details of the proposed custom model are provided in the subsequent sections.
\begin{figure}[b]
    \centering
    \includegraphics[width=\linewidth]{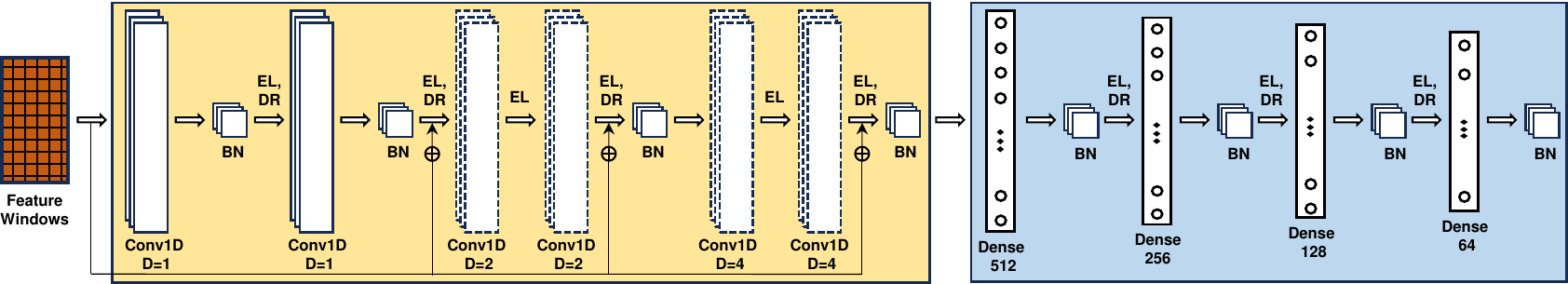}
    \caption{Proposed EEdGeNet model architecture integrating a Temporal Convolutional Block (TCB) and a Dense Transformation Block (DTB). The TCB (highlighted in yellow) captures temporal dependencies via dilated convolutions, while the DTB (highlighted in blue) performs hierarchical non-linear feature transformations. Abbreviations: BN—Batch Normalization, DR—Dropout layer, EL—ELU activation, and $D$—dilation rate.}
    \label{fig:model}
\end{figure}
Figure \ref{fig:model} shows the architecture of the proposed model, comprising a Temporal Convolutional Block (TCB) and a Dense Transformation Block (DTB). The TCB captures temporal dependencies using dilated convolutions, while the DTB performs non-linear transformations via stacked dense layers.

\subsubsection*{Pre-processing and input representation}
In this work, we utilize windows of EEG signals processed through a feature extraction pipeline that computes a predefined set of time-domain and frequency-domain features. The input to the model is a windowed segment \( X_i \in \mathbb{R}^{C \times F} \), where \( C = 32 \) denotes the number of EEG channels and \( F \) represents the number of computed features per channel. The model is designed to map each windowed segment \( X_i \) to its corresponding class label \( y_i \), given a labeled dataset \( \mathcal{S} = \{(X_i, y_i)\}_{i=1}^{m} \), where \( y_i \in \{1, \ldots, n\} \) and \( n = 27 \) denotes the total number of classes in our task.

\subsubsection*{Temporal convolutional block}
To extract spatial and temporal relationships among EEG channels, a TCN is integrated into the model architecture \cite{bai2018empirical}. This block comprises an initial two-layer causal convolutional sub-block with Batch Normalization (BN), Exponential Linear Unit (ELU) activation, and dropout, followed by two dilated residual sub-blocks designed to capture local dependencies and feature interactions within each channel. The input to this block is a windowed segment \( X_i \in \mathbb{R}^{C \times F} \), where \( C \) denotes the number of EEG channels and \( F \) represents the number of computed features per channel.

The input to the TCN block is denoted as \( Z^{(0)} \), and its transformation through the block is described in equation \ref{eq:tcb}.

\begin{equation}
\label{eq:tcb}
\begin{aligned}
Z^{(1)} &= \mathrm{Dropout}\left(\phi\left(\mathrm{BN}\left(\mathrm{Conv1D}_{k=10, d=1}\left(
        \mathrm{Dropout}\left(\phi(\mathrm{BN}(\mathrm{Conv1D}_{k=10, d=1}(Z^{(0)})))\right)\right)\right)\right)\right), \\
Z^{(2)} &= \mathrm{Dropout}\left(\phi\left(\mathrm{BN}\left(\mathrm{Conv1D}_{k=3, d=d_j}\left(
        \phi\left(\mathrm{BN}\left(\mathrm{Conv1D}_{k=3, d=d_i}(Z^{(j)})\right)\right)\right)\right)\right)\right), \quad j = 1, 2;\; d_i = 2,\; d_j = 4,
\end{aligned}
\end{equation}

where \( Z^{(1)} \) is the output of the initial causal convolutional sub-block, and \( Z^{(2)} \) is the output from the two dilated residual sub-blocks. Here, \( \phi(\cdot) \) denotes the ELU activation function, \( k \) is the kernel size, \( d \) is the dilation rate, and \( \mathrm{BN} \) represents Batch Normalization.

The final output \( Z^{(2)} \) is subsequently flattened and passed to the fully connected layers in the Dense Transformation Block to learn complex, non-linear representations from the extracted features.













\subsubsection*{Dense transformation block}
The flattened representation \( Z^{(2)} \) is passed through a hierarchical network of fully connected layers with progressively decreasing dimensionality to learn high-level discriminative representations for classification. Each dense layer is followed by BN, ELU activation, and dropout, which collectively enhance training stability and reduce the risk of overfitting.

The hidden layers are configured with the following number of neurons: \( 512 \rightarrow 256 \rightarrow 128 \rightarrow 64 \). The transformation at each layer is defined in equation \ref{eq:dtb}:

\begin{equation}
\label{eq:dtb}
h^{(l)} = \text{Dropout} \left( \sigma \left( \text{BN} \left( W^{(l)} h^{(l-1)} + b^{(l)} \right) \right) \right), \quad l = 1, 2, 3, 4
\end{equation}

where \( h^{(0)} = \text{Flatten}(Z^{(2)}) \in \mathbb{R}^{CF} \) denotes the flattened input vector, \( \sigma(\cdot) \) is the ELU activation function, and BN represents Batch Normalization.

The output from the final dense layer is passed through a softmax classification layer to produce the class probability distribution for subsequent label prediction.

Table \ref{tab:hyperparameters} lists the hyperparameter configurations of the Temporal Convolutional Block and Dense Transformation Block used in the proposed EEdGeNet model. Hyperparameters were selected using scikit-learn’s \texttt{GridSearchCV} with 3-fold cross-validation.

\begin{table}[b]
\centering
\caption{Hyperparameter configuration of the Temporal Convolutional Block (TCB) and Dense Transformation Block (DTB) used in the proposed EEdGeNet model.}
\label{tab:hyperparameters}
\begin{adjustbox}{width=0.5\textwidth}
\begin{tabular}{|c|c|c|}
\hline
\textbf{Block} & \textbf{Hyperparameter} & \textbf{Value} \\
\hline
\multirow{4}{*}{\makecell{Temporal\\ Convolutional\\ Block (TCB)}} 
& Dilation Rates & [1, 2, 4] \\
\cline{2-3}
& Kernel Size & 3 \\
\cline{2-3}
& Filters per Conv Layer & 32 \\
\cline{2-3}
& Dropout Rate & 0.3 \\
\hline
\multirow{3}{*}{\makecell{Dense\\ Transformation\\ Block (DTB)}} 
& Hidden Layers & [512, 256, 128, 64] \\
\cline{2-3}
& Dropout Rate & 0.3 \\
\cline{2-3}
& L2 Regularization & 0.0008 \\
\hline
\end{tabular}
\end{adjustbox}
\end{table}

\subsection*{Experimental Setup}
\subsection*{Model training}
The EEG dataset was used to train the EEdGeNet model locally on the personal machine. Adam optimizer was used for training the custom model along with a cosine decay learning rate schedule for learning rate reduction over a set time. The initial learning rate was set to \( 3 \times 10^{-4} \) and was set to decay until a minimum learning rate of \( 1 \times 10^{-5} \), decay steps were defined as \( \texttt{decay\_steps} = \left\lfloor \frac{N}{B} \right\rfloor \times E \), where \( N \) is the number of training samples, \( B = 32 \) is the batch size, and \( E = 500 \) is the maximum number of epochs.

To calculate the dissimilarity of the predicted probability distribution from the model and the actual class labels, sparse categorical cross-entropy loss was used. The model was trained with early stopping, monitoring the validation loss with patience of 15 epochs and restored the weights from the epoch with the lowest validation loss. In order to address any possible class imbalance in the training set, class weights were calculated and incorporated during training to assign higher importance to minority classes. A model checkpoint callback saved the weights corresponding to the highest validation accuracy observed during training.

\subsection*{Local machine-based inference}
Inference was performed on the test dataset using a batch size of 1, enabling character-by-character prediction. The entire pipeline—including preprocessing, feature extraction, and model evaluation—was executed in real time. The experiments were conducted on a personal machine equipped with an Intel Core Ultra 9-275HX processor (24 cores, 2.7 GHz), 32 GB of RAM, and an NVIDIA GeForce RTX 5060 Laptop GPU.
\subsection*{Inference on an edge device}
To demonstrate the portability of the proposed real-time EEG-based handwritten character recognition system, inference was also conducted on an edge device. Specifically, the NVIDIA Jetson TX2 platform was used to evaluate the trained models, performing character-by-character prediction with a batch size of 1. The Jetson TX2 features a 256-core NVIDIA Pascal™ GPU, a dual-core NVIDIA Denver 2 64-bit CPU, and a quad-core ARM® Cortex®-A57 MPCore. It is equipped with 8 GB of 128-bit LPDDR4 memory (1866 MHz, 59.7 GB/s bandwidth) and 32 GB of eMMC 5.1 storage. The device operates within a configurable power envelope of 7.5W to 15W.

\textcolor{black}{To measure energy consumption during inference, we employed NVIDIA’s tegrastats utility, which provides real-time board-level power readings via the VDD\_IN rail. Power values were logged at 100 ms intervals while the inference script was executed, and the average power draw was computed over the entire inference duration. Multiplying this average by the elapsed runtime yielded the total energy, which was then normalized by the number of processed EEG windows
to report the per-character inference energy cost.}

\section*{Experimental Results}
\begin{table*}[b]
\small
\centering
\caption{Performance comparison of model configurations applied to raw and feature-extracted EEG signals for imagined handwritten character recognition. Among the evaluated architectures, the combination of TCN and MLP (EEdGeNet) on 85 extracted features achieves the highest classification accuracy and lowest per-character latency.}
\label{tab:eeg_model_summary}
\begin{adjustbox}{width=\textwidth}
\begin{tabular}{@{}c c c c c c c c@{}}
\toprule
\textbf{Model} & \makecell{\textbf{Test Accuracy (\%)}} & \textbf{95\% CI (\%)} & \textbf{Precision (\%) } & \textbf{Recall (\%)} & \textbf{F1-Score (\%)} & \makecell{\textbf{Inference Time/Character} \\ \textbf{(Local Machine)}} & \makecell{\textbf{Inference Time/Character} \\ \textbf{(NVIDIA Jetson TX2)}} \\
\midrule
MLP (Raw dataset) 
& 10.26 $\pm$ 0.13 
& [10.13, 10.40] 
& 10.87 $\pm$ 0.13 
& 10.26 $\pm$ 0.17 
& 10.06 $\pm$ 0.09 
& 6.70 ms & N/A \\
\midrule
TCN (Raw dataset) 
& 29.10 $\pm$ 1.03 
& [27.83, 30.38] 
& 30.24 $\pm$ 0.96 
& 29.10 $\pm$ 1.28 
& 29.15 $\pm$ 1.25 
& 6.70 ms & N/A \\
\midrule
EEdGeNet (Raw dataset) 
& 43.63 $\pm$ 2.77 
& [43.19, 50.08] 
& 47.46 $\pm$ 3.20 
& 46.63 $\pm$ 3.44 
& 46.56 $\pm$ 3.41 
& 10.41 ms & N/A \\
\midrule
MLP (85 Features) 
& 73.82 $\pm$ 0.58 
& [72.06, 75.59] 
& 74.07 $\pm$ 0.55 
& 73.82 $\pm$ 0.58 
& 73.88 $\pm$ 0.58 
& 40.98 ms & N/A \\
\midrule
TCN (85 Features) 
& 76.95 $\pm$ 0.10 
& [76.64, 77.26] 
& 77.01 $\pm$ 0.08 
& 76.95 $\pm$ 0.10 
& 76.95 $\pm$ 0.09 
& 40.98 ms & N/A\\
\midrule
\textbf{EEdGeNet (85 Features)} 
& \textbf{89.83 $\pm$ 0.19} 
& \textbf{[89.56, 90.09]} 
& \textbf{89.87 $\pm$ 0.18} 
& \textbf{89.83 $\pm$ 0.19} 
& \textbf{89.83 $\pm$ 0.19} 
& \textbf{41.25 ms} & \textbf{914.18 ms} \\
\bottomrule
\end{tabular}
\end{adjustbox}
\end{table*}

\subsection*{Towards an optimal architecture for imagined handwriting from EEG}
Table \ref{tab:eeg_model_summary} presents a comparative evaluation of various model configurations for imagined handwritten character classification from EEG signals, considering both raw and feature-extracted inputs. A four-layer multilayer perceptron (MLP) with hidden dimensions of 512, 256, 128, and 64, when applied directly to the raw EEG signals, yields a test accuracy of 10.26\%~$\pm$~0.13\%, with limited discriminative capability. The use of a TCN on the raw data offers modest improvement, reaching 29.10\%~$\pm$~1.03\%, consistent with prior reports on TCNs outperforming recurrent models like LSTM and GRU for sequence modeling tasks~\cite{bai2018empirical}. To enhance performance, we extended the architecture by appending our proposed MLP module to the TCN backbone. This hybrid TCN-MLP model (EEdGeNet) achieves 43.63\%~$\pm$~2.77\% accuracy on the raw dataset. While models using raw signals maintain low per-character inference latency on the local machine (approximately 7-11ms), their classification performance remains suboptimal.

To overcome this limitation, we extracted a compact set of 85 features were then extracted, covering the time and frequency domains, along with graphical descriptors, as described in the \textit{Feature Extraction} section. When evaluated independently, MLP and TCN models trained on these extracted features yield accuracies of 73.82\%~$\pm$~0.58\% and 72.62\%~$\pm$~0.10\%, respectively---demonstrating the utility of domain-informed representations. The best overall performance is obtained with the feature-based EEdGeNet model, which achieves 89.83\%~$\pm$~0.19\% test accuracy, accompanied by high precision, recall, and F1-score metrics. The corresponding inference latency per-character on a local machine is 41.25 ms, while inference on the NVIDIA Jetson TX2 requires 914.18 ms per-character, as the computation involves additional layers and embedded resource constraints. In subsequent analyses, we address this bottleneck by reducing the feature set to optimize on-device efficiency without significantly sacrificing classification performance.
\subsection*{Comparative analysis of time domain, frequency domain, and graphical features}  
\begin{table*}[t]
\small
\centering
\caption{Comparison of results across selected time-domain, frequency-domain, and graphical features evaluated using the EEdGeNet model. Frequency-domain features yield lower test accuracy but substantially reduced per-character latency during inference. In contrast, time-domain features achieve higher accuracy across the full feature set but incur greater latency. Graphical features provide a balance between accuracy and efficiency, while combining all 85 features delivers the best overall performance for per-character prediction on the NVIDIA Jetson TX2.}
\label{tab:time_frequency_compare}
\begin{adjustbox}{width=\textwidth}
\begin{tabular}{@{}c c c c c c c c@{}}
\toprule
\textbf{Feature Set} 
& \makecell{\textbf{Test Accuracy (\%)}} 
& \textbf{95\% CI (\%)} 
& \textbf{Precision (\%)} 
& \textbf{Recall (\%)} 
& \textbf{F1-Score (\%)} 
& \makecell{\textbf{Inference Time/Character} \\ \textbf{(Local Machine)}} 
& \makecell{\textbf{Inference Time/Character} \\ \textbf{(NVIDIA Jetson TX2)}} \\
\midrule
\makecell{\textbf{Time Domain} \\ \textbf{Features (12)}} 
& 85.40 $\pm$ 0.88
& [84.48, 86.32]
& 86.18 $\pm$ 0.20
& 85.89 $\pm$ 0.19
& 85.88 $\pm$ 0.19
& 36.93 ms 
& 792.73 ms \\
\midrule
\makecell{\textbf{Frequency Domain} \\ \textbf{Features (8)}} 
& 72.18 $\pm$ 0.16
& [71.69, 72.67]
& 72.39 $\pm$ 0.19
& 72.18 $\pm$ 0.16
& 72.22 $\pm$ 0.17
& 7.24 ms 
& 199.52 ms \\
\midrule
\makecell{\textbf{Graphical} \\ \textbf{Features (65)}} 
& 83.26 $\pm$ 1.29
& [81.46, 85.06]
& 83.52 $\pm$ 1.29
& 83.26 $\pm$ 1.29
& 83.26 $\pm$ 1.31
& 16.94 ms 
& 229.84 ms \\
\midrule
\makecell{\textbf{All Features (85)}} 
& 89.83 $\pm$ 0.19
& [89.56, 90.09]
& 89.87 $\pm$ 0.18
& 89.83 $\pm$ 0.19
& 89.83 $\pm$ 0.19
& 41.25 ms 
& 914.18 ms\\
\bottomrule
\end{tabular}
\end{adjustbox}
\end{table*}
Table \ref{tab:time_frequency_compare} presents a comparative evaluation of time-domain, frequency-domain, and graphical feature sets for imagined handwriting classification from EEG signals using the proposed EEdGeNet model. Time-domain features achieved the highest standalone accuracy ($85.40\% \pm 0.88\%$) but incurred relatively high inference latency (36.93 ms per-character on a local machine and 792.73 ms on the NVIDIA Jetson TX2). By contrast, frequency-domain features offered the lowest latency (7.24 ms on local machine and 199.52 ms on Jetson TX2) but correspondingly yielded the lowest accuracy ($72.18\% \pm 0.16\%$). Graphical features provided an intermediate trade-off, attaining $83.26\% \pm 1.29\%$ accuracy with moderate latency (16.94 ms locally and 229.84 ms on Jetson TX2). When all 85 features were combined, overall accuracy was maximized ($89.83\% \pm 0.19\%$), albeit at the cost of increased latency (41.25 ms locally and 914.18 ms on Jetson TX2). Although the full feature set achieves the highest accuracy, the accompanying latency underscores the need for feature selection, motivating the search for a minimal subset that preserves accuracy while substantially reducing inference time. This direction is pursued in the following subsection.

\subsection*{Impact of feature selection on model performance}
\begin{figure}[h]
    \centering
    \includegraphics[width=\linewidth]{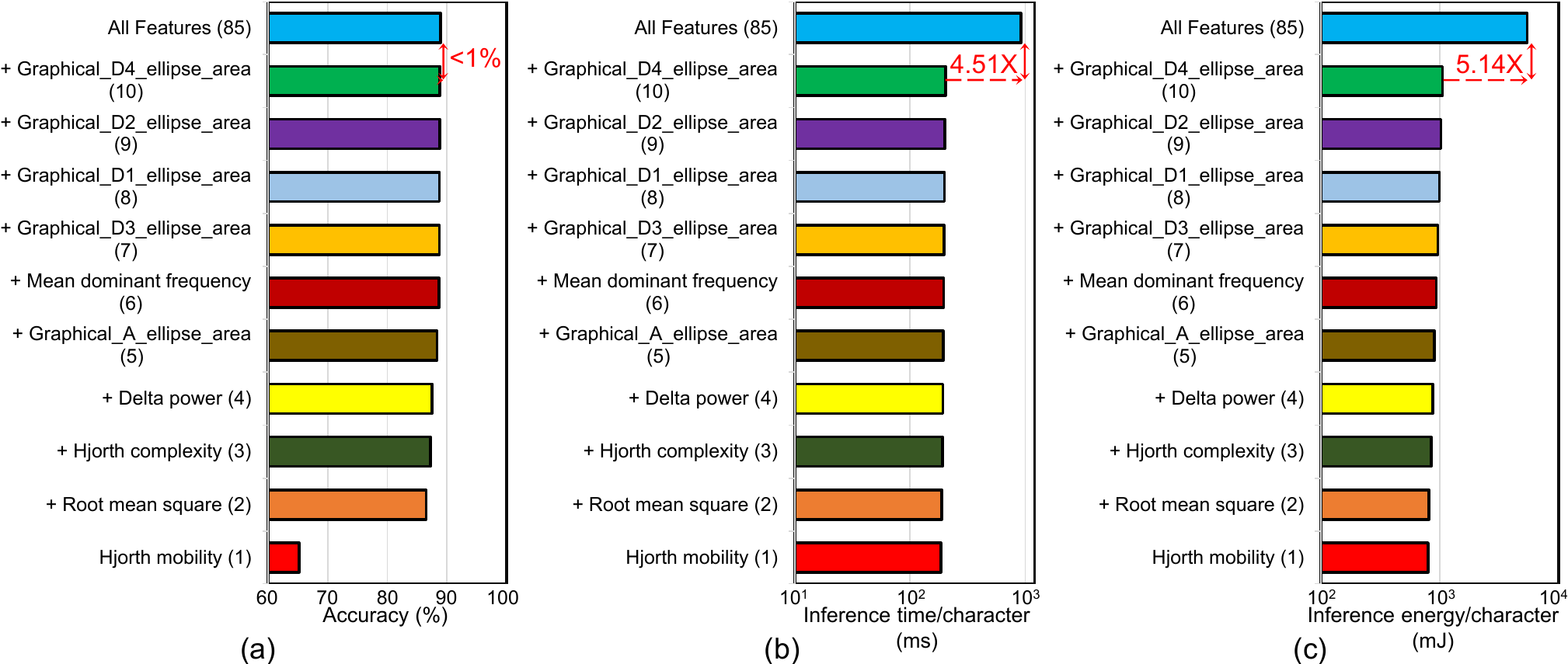}
    \caption{\black{Impact of feature selection using Pearson correlation coefficients on classification accuracy, inference latency, and energy consumption on the NVIDIA Jetson TX2. A compact feature set comprising Hjorth mobility, root mean square, Hjorth complexity, delta power, graphical\_A ellipse area, mean dominant frequency, graphical\_D3 ellipse area, graphical\_D1 ellipse area, graphical\_D2 ellipse area, and graphical\_D4 ellipse area results in (a) a $\sim$1\% reduction in accuracy relative to the full 85-feature set evaluated on the proposed EEdGeNet model, (b) a 4.51$\times$ improvement in per-character inference time, and (c) a 5.14$\times$ reduction in per-character inference energy. This trade-off highlights an efficient balance between computational cost and predictive performance, well-suited for edge deployment on the Jetson TX2.}}
    \label{fig:feature_selection}
\end{figure}
\textcolor{black}{To identify the most informative features, we conducted a systematic feature selection analysis based on Pearson correlation coefficients computed across all 32 EEG channels for each character and participant. In a representative case involving one participant and one character, some of the negatively correlated feature pairs were: (Hjorth mobility vs. Hjorth complexity), (Hjorth mobility vs. delta power), (RMS vs. Hjorth mobility), (kurtosis vs. Shannon entropy), (Hjorth mobility vs. graphical\_A\_ellipse area), (Hjorth mobility vs. graphical\_A\_Std\_X-Axis), (Hjorth mobility vs. graphical\_A\_Std\_Y-Axis), (Hjorth mobility vs. graphical\_A\_Mean\_Distance\_Origin), (Hjorth mobility vs. mean dominant frequency), (RMS vs. mean dominant frequency), (graphical\_D1\_ellipse area vs. graphical\_D1\_covariance of x-y coordinates), (graphical\_D2\_ellipse area vs. graphical\_D2\_covariance of x-y coordinates), (delta power vs. mean dominant frequency), (delta power vs. spectral entropy), and (Hjorth complexity vs. gamma power).
Conversely, in the same participant--character instance, some of the most positively correlated feature pairs were: (RMS vs. Variance), (Hjorth mobility vs. spectral entropy), (Hjorth mobility vs. gamma power), (graphical\_A\_ellipse area vs. graphical\_A\_Std\_X-Axis), (graphical\_A\_ellipse area vs. graphical\_A\_Std\_Y-Axis), and (graphical\_A\_ellipse area vs. graphical\_A\_Mean\_Distance\_Origin).}

\black{These associations revealed redundancy among several features. For example, RMS and Variance were strongly correlated, suggesting that Variance may be omitted without substantial information loss. Similarly, spectral entropy and gamma power exhibited strong correlations with Hjorth mobility, indicating redundancy. Likewise, graphical\_A\_ellipse area could substitute for graphical\_A\_Std\_X-Axis, graphical\_Std\_Y-Axis, and graphical\_A\_Mean\_Distance\_Origin.
Extending this analysis across all remaining feature pairs, we identified a consistent set of features that remained distinct: Hjorth mobility, root mean square (RMS), Hjorth complexity, delta power, graphical\_A\_ellipse area, mean dominant frequency, graphical\_D3\_ellipse area, graphical\_D1\_ellipse area, graphical\_D2\_ellipse area, and graphical\_D4\_ellipse area. Importantly, this ordering was preserved across all 27 character classes and 15 participants, highlighting the robustness and generalizability of our feature selection procedure.}

\black{Figure~\ref{fig:feature_selection}a--c illustrates the trade-offs among selected feature subsets in terms of classification accuracy on the proposed EEdGeNet model, along with per-character inference latency and energy consumption measured on the NVIDIA Jetson TX2. Feature subsets were derived using Pearson correlation-based ranking, as described in the \textit{Feature Selection} section. Employing only the most discriminative feature—Hjorth mobility—achieves an accuracy of 65.18\%, with an inference latency of 184.71\,ms and energy consumption of 796.2\,mJ per-character. Incremental inclusion of root mean square, Hjorth complexity, delta power, graphical\_A ellipse area, mean dominant frequency, graphical\_D3 ellipse area, graphical\_D1 ellipse area, graphical\_D2 ellipse area, and graphical\_D4 ellipse area progressively enhances performance, ultimately achieving $88.84\% \pm 0.09\%$ accuracy, 202.62\,ms latency, and 1052.42\,mJ energy usage. Notably, this 10-feature configuration delivers accuracy within $\sim$1\% of the full 85-feature set ($89.83\% \pm 0.19\%$), while reducing inference time and energy consumption by 4.51$\times$ and 5.14$\times$, respectively. These results highlight an effective trade-off between computational efficiency and predictive performance, making the system well-suited for real-time inference on resource-constrained edge devices.}

\subsection*{Comparative Evaluation with Existing Methods}
\subsubsection*{\black{Ablation study of preprocessing, feature, and classifier variants}}  
\black{We performed an extensive ablation analysis to disentangle the contributions of preprocessing, feature extraction, and classifier choices within our proposed real-time EEG-based handwriting recognition pipeline. The comparative results, summarized in Table~\ref{tab:unified_ablation_updated}, demonstrate how each design choice influences classification accuracy, inference latency, and energy efficiency on the NVIDIA Jetson TX2.}  

\black{Replacing the simplified ASR with conventional denoising methods yielded mixed outcomes. Standard ASR substantially reduced accuracy ($80.80 \pm 0.15\%$), while ICA collapsed performance ($12.03 \pm 0.15\%$), underscoring their unsuitability for imagined handwriting decoding. Decomposition-based approaches such as VMD \cite{sadiq2022motor} and EWT \cite{sadiq2019motor} achieved accuracies of $89.15 \pm 0.01\%$ and $87.06 \pm 0.34\%$, respectively, but incurred markedly higher inference times (2026--2738 ms) and energy costs (6807--9200 mJ). MSPCA provided competitive accuracy ($87.12 \pm 1.46\%$) with the lowest latency (828.93 ms), but still underperformed relative to simplified ASR in terms of overall trade-off.}  

\black{We next investigated alternative feature representations. While reconstructed phase-space descriptors (13 features) \cite{akbari2021depression}, fast fractional Fourier descriptors (26 features) \cite{sadiq2024fast}, and Poincaré–DWT graphical features (65 features) \cite{akbari2023recognizing }reduced inference latency (200--400 ms) and per-character energy consumption, their accuracies ranged only from $73.55 \pm 0.10\%$ to $83.26 \pm 1.29\%$, falling well below the baseline. These results indicate that aggressive feature compaction or reliance on handcrafted graphical descriptors improves computational efficiency on the Jetson TX2 but sacrifices predictive performance.}  

\black{Finally, we evaluated alternative classifiers. A 1D CNN–LSTM trained on the full handcrafted feature set achieved modest accuracy ($77.55 \pm 0.88\%$) and latency (969.24 ms). In contrast, an image-based ShuffleNet pretrained on ImageNet \cite{sadiq2022exploiting} was found to be impractical for real-time deployment: generating canonical wavelet transform (CWT) images from raw EEG required 438.20 ms/character on our local machine, compared to only 41.25 ms/character for our handcrafted feature pipeline. Thus, TF-image generation is approximately 10.62$\times$ slower than direct feature extraction. Consequently, TF-image pipelines are unsuitable for per-character, real-time inference in edge-deployed BCI applications.}  

\black{Taken together, these ablations underscore that the proposed pipeline—simplified ASR followed by 85 physiologically grounded handcrafted features and the EEdGeNet model—delivers the most favorable balance between accuracy, latency, and energy efficiency. This configuration consistently outperformed all alternative pipelines, establishing it as the most effective architecture for real-time imagined handwriting recognition on edge devices.} 

\begin{table*}[t]
\small
\centering
\caption{Ablation study of the proposed real-time EEG-based handwriting recognition pipeline, spanning preprocessing, feature extraction, and classifier variants. Alternatives such as MSPCA, reconstructed phase-space features (13), fast fractional Fourier descriptors (26), and Poincaré–DWT graphical features (65) reduce inference time and energy consumption on the NVIDIA Jetson TX2 but incur notable losses in accuracy. By contrast, the proposed pipeline with simplified ASR, the full 85 handcrafted features, and the EEdGeNet model achieves the highest per-character classification accuracy while maintaining a favorable trade-off between accuracy, latency, and energy efficiency.}
\label{tab:unified_ablation_updated}
\begin{adjustbox}{width=\textwidth}
\begin{tabular}{@{}c c c c c@{}}
\toprule
\textbf{Study} &
\textbf{Ablation Study Pipeline} &
\textbf{Test Accuracy (\%)} &
\makecell{\textbf{Inference Time/Character}\\\textbf{(NVIDIA Jetson TX2)}} &
\makecell{\textbf{Inference Energy/Character}\\\textbf{(NVIDIA Jetson TX2)}} \\
\midrule
\textbf{This work} &
\makecell[l]{\textbf{Raw Data $\rightarrow$ Train/Test Split $\rightarrow$ Bandpass (1--50 Hz)}\\
\textbf{$\rightarrow$ Simplified ASR $\rightarrow$ Features (85) $\rightarrow$ EEdGeNet}} &
\textbf{\textcolor{blue}{89.83 $\pm$ 0.19}} &
\textbf{914.18 ms} &
\textbf{5406.84 mJ} \\
\midrule
1 &
\makecell[l]{Raw Data $\rightarrow$ Train/Test Split $\rightarrow$ Bandpass (1--50 Hz)\\
$\rightarrow$ \textit{Standard ASR} $\rightarrow$ Features (85) $\rightarrow$ EEdGeNet} &
80.80 $\pm$ 0.15 &
1501.48 ms &
5051.92 mJ \\
\midrule
2 &
\makecell[l]{Raw Data $\rightarrow$ Train/Test Split $\rightarrow$ Bandpass (1--50 Hz)\\
$\rightarrow$ \textit{ICA} $\rightarrow$ Features (85) $\rightarrow$ EEdGeNet} &
\textcolor{red}{12.03 $\pm$ 0.15} &
\textcolor{red}{5492.51 ms} &
\textcolor{red}{18454.83 mJ} \\
\midrule
3 &
\makecell[l]{Raw Data $\rightarrow$ Train/Test Split $\rightarrow$ Bandpass (1--50 Hz)\\
$\rightarrow$ \textit{VMD} $\rightarrow$ Features (85) $\rightarrow$ EEdGeNet} &
89.15 $\pm$ 0.01 &
2026.02 ms &
6807.42 mJ \\
\midrule
4 &
\makecell[l]{Raw Data $\rightarrow$ Train/Test Split $\rightarrow$ Bandpass (1--50 Hz)\\
$\rightarrow$ \textit{EWT} $\rightarrow$ Features (85) $\rightarrow$ EEdGeNet} &
87.06 $\pm$ 0.34 &
2738.36 ms &
9200.89 mJ \\
\midrule
5 &
\makecell[l]{Raw Data $\rightarrow$ Train/Test Split $\rightarrow$ Bandpass (1--50 Hz)\\
$\rightarrow$ \textit{MSPCA} $\rightarrow$ Features (85) $\rightarrow$ EEdGeNet} &
87.12 $\pm$ 1.46 &
828.93 ms &
5153.42 mJ \\
\midrule
6 &
\makecell[l]{Raw Data $\rightarrow$ Train/Test Split $\rightarrow$ Bandpass (1--50 Hz)\\
$\rightarrow$ \textit{EWT} $\rightarrow$ Features (20) $\rightarrow$ EEdGeNet} &
87.06 $\pm$ 0.34 &
2738.36 ms &
9200.89 mJ \\
\midrule
7 &
\makecell[l]{Raw Data $\rightarrow$ Train/Test Split $\rightarrow$ Bandpass (1--50 Hz)\\
$\rightarrow$ \textit{MSPCA} $\rightarrow$ Features (20) $\rightarrow$ EEdGeNet} &
87.12 $\pm$ 1.46 &
828.93 ms &
5153.42 mJ \\
\midrule
8 &
\makecell[l]{Raw Data $\rightarrow$ Train/Test Split $\rightarrow$ Bandpass (1--50 Hz)\\
$\rightarrow$ Simplified ASR $\rightarrow$ \textit{Reconstructed Phase-Space Features (13)} $\rightarrow$ EEdGeNet} &
82.32 $\pm$ 1.07 &
202.10 ms &
1324.31 mJ \\
\midrule
9 &
\makecell[l]{Raw Data $\rightarrow$ Train/Test Split $\rightarrow$ Bandpass (1--50 Hz)\\
$\rightarrow$ Simplified ASR $\rightarrow$ \textit{Fast Fractional Fourier Descriptors (26)} $\rightarrow$ EEdGeNet} &
\textcolor{red}{73.55 $\pm$ 0.10} &
\textcolor{blue}{200.99 ms} &
\textcolor{blue}{1317.04 mJ} \\
\midrule
10 &
\makecell[l]{Raw Data $\rightarrow$ Train/Test Split $\rightarrow$ Bandpass (1--50 Hz)\\
$\rightarrow$ Simplified ASR $\rightarrow$ \textit{Poincaré--DWT graphical Features (65)} $\rightarrow$ EEdGeNet} &
83.26 $\pm$ 1.29 &
229.84 ms &
1506.09 mJ \\
\midrule
11 &
\makecell[l]{Raw Data $\rightarrow$ Train/Test Split $\rightarrow$ Bandpass (1--50 Hz)\\
$\rightarrow$ Simplified ASR $\rightarrow$ Features (85) $\rightarrow$ \textit{1D CNN--LSTM}} &
77.55 $\pm$ 0.88 &
969.24 ms &
5447.83 mJ \\
\midrule
12 &
\makecell[l]{Raw Data $\rightarrow$ Train/Test Split $\rightarrow$ MSPCA $\rightarrow$ \textit{CWT}\\
$\rightarrow$ \textit{ShuffleNet (pretrained on ImageNet)}} &
N/A &
N/A &
N/A \\
\bottomrule
\end{tabular}
\end{adjustbox}
\end{table*}

\subsubsection*{Performance evaluation of the custom model against state-of-the-art architectures}
\begin{table}[h]
\centering
\caption{Comparative performance analysis of state-of-the-art models and the proposed custom architecture on the raw imagined handwriting EEG dataset and 85 features extracted data. The proposed EEdGeNet model trained on 85 extracted features achieves the highest test accuracy and F1-score, outperforming existing methods across all evaluated metrics.}
\label{tab:sota_comparison}
\begin{adjustbox}{width=0.95\textwidth}
\renewcommand{\arraystretch}{1.2}
\begin{tabular}{|c|c|c|c|c|c|c|c|c|}
\hline
\multirow{2}{*}{\textbf{Model}} &
\multicolumn{3}{c|}{\textbf{Raw Data}} &
\multicolumn{3}{c|}{\textbf{Features (85)}} \\
\cline{2-7}
& \textbf{Test Acc. (\%)} & \textbf{95\% CI} & \multicolumn{1}{c|}{\makecell{\textbf{Precision (P) (\%)}\\\textbf{Recall (R) (\%)}\\\textbf{F1-Score (F) (\%)}}}
& \textbf{Test Acc. (\%)} & \textbf{95\% CI} & \multicolumn{1}{c|}{\makecell{\textbf{Precision (P) (\%)}\\\textbf{Recall (R) (\%)}\\\textbf{F1-Score (F) (\%)}}}  \\
\hline
\multirow{2}{*}{DeepConvNet \cite{schirrmeister2017deep}} 
& 35.89 $\pm$ 1.42 & [34.13, 37.64] & \makecell{P: 40.64 $\pm$ 1.30 \\ R: 35.89 $\pm$ 1.42 \\ F: 34.55 $\pm$ 1.45} 
& 
52.71 $\pm$ 1.64 & [48.64, 56.78] & \makecell{P: 54.29 $\pm$ 1.22 \\ R: 52.71 $\pm$ 1.64 \\ F: 52.58 $\pm$ 1.66}  \\
\hline
\multirow{2}{*}{ShallowConvNet \cite{schirrmeister2017deep}} 
& 43.49 $\pm$ 0.57 & [42.78, 44.20] & \makecell{P: 43.87 $\pm$ 0.72 \\ R: 43.49 $\pm$ 0.57 \\ F: 43.24 $\pm$ 0.64} 
& 
41.10 $\pm$ 0.18 & [40.65, 41.55] & \makecell{P: 41.43 $\pm$ 0.28 \\ R: 41.10 $\pm$ 0.18 \\ F: 40.98 $\pm$ 0.24} \\
\hline
\multirow{2}{*}{EEGNet \cite{lawhern2018eegnet}} 
& 26.17 $\pm$ 1.47 & [24.13, 28.22] & \makecell{P: 29.99 $\pm$ 1.05 \\ R: 26.17 $\pm$ 1.47 \\ F: 24.99 $\pm$ 1.55} 
& 
43.37 $\pm$ 0.74 & [41.12, 45.62] & \makecell{P: 44.42 $\pm$ 0.59 \\ R: 43.37 $\pm$ 0.74 \\ F: 42.76 $\pm$ 0.77}  \\
\hline
\multirow{2}{*}{TCNet \cite{bai2018empirical}} 
& 29.10 $\pm$ 1.03 & [27.83, 30.38] & \makecell{P: 30.24 $\pm$ 0.96 \\ R: 29.10 $\pm$ 1.28 \\ F: 29.15 $\pm$ 1.25} 
& 
76.95 $\pm$ 0.10 & [76.64, 77.26] & \makecell{P: 77.01 $\pm$ 0.08 \\ R: 76.95 $\pm$ 0.10 \\ F: 76.95 $\pm$ 0.09}  \\
\hline
\multirow{2}{*}{EEGTCNet \cite{ingolfsson2020eeg}} 
& 28.92 $\pm$ 3.48 & [24.09, 33.74] & \makecell{P: 32.22 $\pm$ 3.33 \\ R: 28.92 $\pm$ 3.48 \\ F: 28.05 $\pm$ 3.70} 
& 
57.35 $\pm$ 1.02 & [54.26, 60.45] & \makecell{P: 58.14 $\pm$ 1.34 \\ R: 57.35 $\pm$ 1.02 \\ F: 57.26 $\pm$ 1.07}  \\
\hline
\multirow{2}{*}{TCNet-Fusion \cite{musallam2021electroencephalography}} 
& 27.65 $\pm$ 1.11 & [26.28, 29.03] & \makecell{P: 28.97 $\pm$ 0.78 \\ R: 27.65 $\pm$ 1.11 \\ F: 27.52 $\pm$ 1.09} 
& 
62.54 $\pm$ 0.17 & [62.11, 62.96] & \makecell{P: 62.60 $\pm$ 0.13 \\ R: 62.54 $\pm$ 0.17 \\ F: 62.49 $\pm$ 0.16}  \\
\hline
\multirow{2}{*}{Attention Multi Branch CNN \cite{altuwaijri2022multi}} 
& 51.42 $\pm$ 1.14 & [50.01, 52.83] & \makecell{P: 52.38 $\pm$ 1.32 \\ R: 51.42 $\pm$ 1.14 \\ F: 51.52 $\pm$ 1.18} 
& 
62.80 $\pm$ 0.52 & [61.52, 64.08] & \makecell{P: 62.91 $\pm$ 0.52 \\ R: 62.80 $\pm$ 0.52 \\ F: 62.77 $\pm$ 0.52}  \\
\hline
\multirow{2}{*}{ATCNet \cite{altaheri2022physics}} 
& 34.53 $\pm$ 2.66 & [30.85, 38.22] & \makecell{P: 37.44 $\pm$ 3.34 \\ R: 34.53 $\pm$ 2.66 \\ F: 33.49 $\pm$ 2.48} 
& 
58.64 $\pm$ 1.28 & [54.73, 62.54] & \makecell{P: 59.65 $\pm$ 1.19 \\ R: 58.64 $\pm$ 1.28 \\ F: 58.34 $\pm$ 1.38}  \\
\hline
\multirow{2}{*}{\textbf{EEdGeNet (This work)}} 
& \textbf{43.63 $\pm$ 2.77} & \textbf{[43.19, 50.08]} & \makecell{\textbf{P: 47.46 $\pm$ 3.20} \\ \textbf{R: 46.63 $\pm$ 3.44} \\ \textbf{F: 46.56 $\pm$ 3.41}} 
& 
\textbf{89.83 $\pm$ 0.19} & \textbf{[89.56, 90.09]} & \makecell{\textbf{P: 89.87 $\pm$ 0.18} \\ \textbf{R: 89.83 $\pm$ 0.19} \\ \textbf{F: 89.83 $\pm$ 0.19}}  \\
\hline
\end{tabular}
\end{adjustbox}
\end{table}
We conducted a comparative evaluation of our proposed EEdGeNet model against widely used EEG decoding architectures, including DeepConvNet \cite{schirrmeister2017deep}, ShallowConvNet \cite{schirrmeister2017deep}, EEGNet \cite{lawhern2018eegnet}, TCNet \cite{bai2018empirical}, EEGTCNet \cite{ingolfsson2020eeg}, TCNet-Fusion \cite{musallam2021electroencephalography}, Attention Multi-Branch CNN \cite{altuwaijri2022multi}, and ATCNet \cite{altaheri2022physics}. The comparative results, summarized in Table~\ref{tab:sota_comparison}, highlight the performance of these models on both raw EEG signals and feature-extracted data. 

On raw EEG data, accuracies ranged from $26.17 \pm 1.47\%$ (EEGNet) to $51.42 \pm 1.14\%$ (Attention Multi-Branch CNN), reflecting the inherent difficulty of decoding raw EEG recordings. Although our EEdGeNet model achieved $43.63 \pm 2.77\%$ under this setting, none of the evaluated architectures—including our own—delivered performance sufficient for reliable classification of imagined handwritten characters directly from raw EEG.  

\black{When trained on the curated set of 85 extracted features, however, performance improved across nearly all models (with the exception of ShallowConvNet), though the degree of improvement varied. For example, EEGNet reached $43.37 \pm 0.74\%$, while TCNet achieved $76.95 \pm 0.10\%$, illustrating the benefit of incorporating meaningful features. Nevertheless, existing architectures remained limited in their ability to exploit the structured feature space effectively.  
By contrast, our proposed EEdGeNet model, explicitly designed for feature-based inputs, achieved the highest test accuracy ($89.83 \pm 0.19\%$) and consistently outperformed all baselines across precision, recall, and F1-score. Unlike state-of-the-art EEG architectures, which are optimized for raw time-series data where convolutional and pooling layers target oscillatory dynamics, EEdGeNet preserves feature semantics and emphasizes inter-channel dependencies. Its first stage employs a Temporal Convolutional Network (TCN) block to capture temporal relationships across features, followed by a Dense Transformation Block that leverages a deep Multilayer Perceptron (MLP) to learn non-linear mappings and progressively refine discriminative representations. This architecture enables alignment with the structure of extracted EEG features, promoting subject-generalizable learning and superior decoding accuracy.}  

Overall, these findings establish EEdGeNet as a feature-aware architecture that not only achieves state-of-the-art accuracy but also delivers stable and generalizable performance, setting a new benchmark for EEG-based imagined handwriting recognition.

\subsubsection*{Comparison of the proposed framework with existing approaches}
\begin{table*}[t]
\centering
\caption{Comparative analysis of recent imagined motor and speech EEG decoding methods with our proposed work. Our approach represents the first demonstration of real-time, high-accuracy decoding of imagined handwriting from non-invasive EEG signals, implemented on a portable edge device.}
\label{tab:eeg_comparison}
\resizebox{\textwidth}{!}{%
\begin{tabular}{|C{3.5cm}|C{3cm}|C{3.8cm}|C{3.5cm}|C{3.5cm}|C{1.8cm}|C{1.8cm}|C{1.5cm}|}
\hline
\textbf{Study} & \textbf{Application} & \textbf{Dataset Description} & \textbf{Model Type} & \textbf{Classification Accuracy} & \textbf{Real-Time Compatible} & \textbf{Edge Inference} & \textbf{Latency} \\
\hline
Sadiq et al. (2019) \cite{sadiq2019motor} & Motor imagery classification & BCI Competition III (IVa, IVb, IVc), 118 channels, 5 subjects & MEWT + MSPCA + LS-SVM/MLP/LMT & 98\% (subject-dependent), 91.4\% (subject-independent) & No & No & N/A \\
\hline
Ullah et al. (2021) \cite{ullah2021imagined} & Imagined character recognition & EEGMMIDB, BCI IV-2a, 10 subjects, 26 characters & CNN + Morlet wavelet & 99.45\% (4-class), 95.2\% (custom - 26 characters) & No & No & N/A \\ 
\hline
Jerrin et al. (2021) \cite{panachakel2021decoding} & Imagined speech classification & ASU dataset, 64-channel EEG & Transfer learning (ResNet50) & 92.5\% (2-long words),
80\% (3-short words),
79.7\% (3-vowels),
91.4\% (2-short/long words) & No & No & N/A \\
\hline
Altaheri et al. (2022) \cite{altaheri2022physics} & Motor imagery classification & BCI Competition IV 2a & ATCNet (attention + TCN + physics-informed) & 85.38\% (subject - dependent), 70.97\% (subject - independent), (4-class) & No & No & N/A \\
\hline
Xie et al. (2022) \cite{xie2022transformer} & Motor imagery & BCI datasets (raw EEG) & Transformer with spatial-temporal attention & 83.31\% (2-class), 74.44\% (3-class), 64.22\% (4-class) & No & No & N/A \\
\hline
Cho et al. (2022) \cite{cho2021neurograsp} & Grasp classification from motor imagery & 12 participants, 20 channels, 4 grasp types & CNN-BLSTM + Siamese networks & 86\% (offline), 79\% (online) - 2-class; 68\% (offline), 65\% (online) (4-class) & Yes & No & Not reported \\
\hline
Khademi et al. (2022) \cite{khademi2022transfer} & Motor imagery classification & BCI Competition IV 2a & Hybrid CNN-LSTM (Inception-v3, ResNet-50) & Up to 92\% (4-class) & No & No & N/A \\
\hline
Yu et al. (2023) \cite{yu2021new} & Motor and mental imagery detection & BCI Competition datasets, 118 channels, multiple subjects & Hybrid framework (signal decomposition + feature selection + ML classifiers) & Up to 98.5\% (varies with task/dataset) & No & No & N/A \\
\hline
Abdulghani et al. (2023) \cite{abdulghani2023imagined} & Imagined speech recognition & 8-channel EEG & LSTM + wavelet scattering & 92.5\% (4-commands) & No & No & N/A \\
\hline
Alonso Vázquez et al. (2023) \cite{alonso2023eeg} & Spoken word recognition & 32-channel EEG, 28 subjects, 5 Spanish words & SVM, EEGNet  & 91.04\% (binary), 81.2\% (5-words) & No & No & N/A \\
\hline
Amrani et al. (2024) \cite{amrani2024deep} & EEG-to-text decoding & ZuCo v1.0 and v2.0 & EEG encoder + BART + GPT-4 & BLEU-1: 42.75\%, ROUGE-1-F: 33.28\%, BERTScore-F: 53.86\% & No & No & N/A \\
\hline
\textcolor{black}{\textbf{This work}} & \textcolor{black}{\textbf{Imagined handwriting decoding}} & \textcolor{black}{\textbf{32-channel EEG, 15 subjects}} & \textcolor{black}{\textbf{TCN-MLP-Softmax}} & \textcolor{black}{\textbf{88.84\% $\pm$ 0.09\% (27-class)}} & \textcolor{black}{\textbf{Yes}} & \textcolor{black}{\textbf{Yes(Jetson TX2)}} & \textcolor{black}{\textbf{202.62 ms}} \\
\hline
\end{tabular}%
}
\end{table*}
Table \ref{tab:eeg_comparison} presents a comparative analysis of recent studies on imagined motor and speech EEG decoding alongside our proposed approach. Our method constitutes the first demonstration of real-time, high-accuracy decoding of imagined handwriting from non-invasive EEG signals on a portable edge device. It exhibits strong generalizability across multiple participants and achieves low-latency performance, making it suitable for practical real-time assistive communication applications.
\textcolor{black}{Moreover, while prior works such as Skomrock et al. \cite{skomrock2018characterization} and Willett et al. \cite{willett2021high} reported real-time decoding delays of about 821 ms and 1000 ms respectively for different use cases with invasive neural recordings, our non-invasive pipeline achieves per-character inference in 202.62 ms with $88.84\% \pm 0.09\%$ accuracy. This establishes our framework as a safer and more practical alternative for real-world assistive BCI applications.}

\section*{\black{Discussion}}  
\begin{figure}[t]
    \centering
    \includegraphics[width=0.8\linewidth]{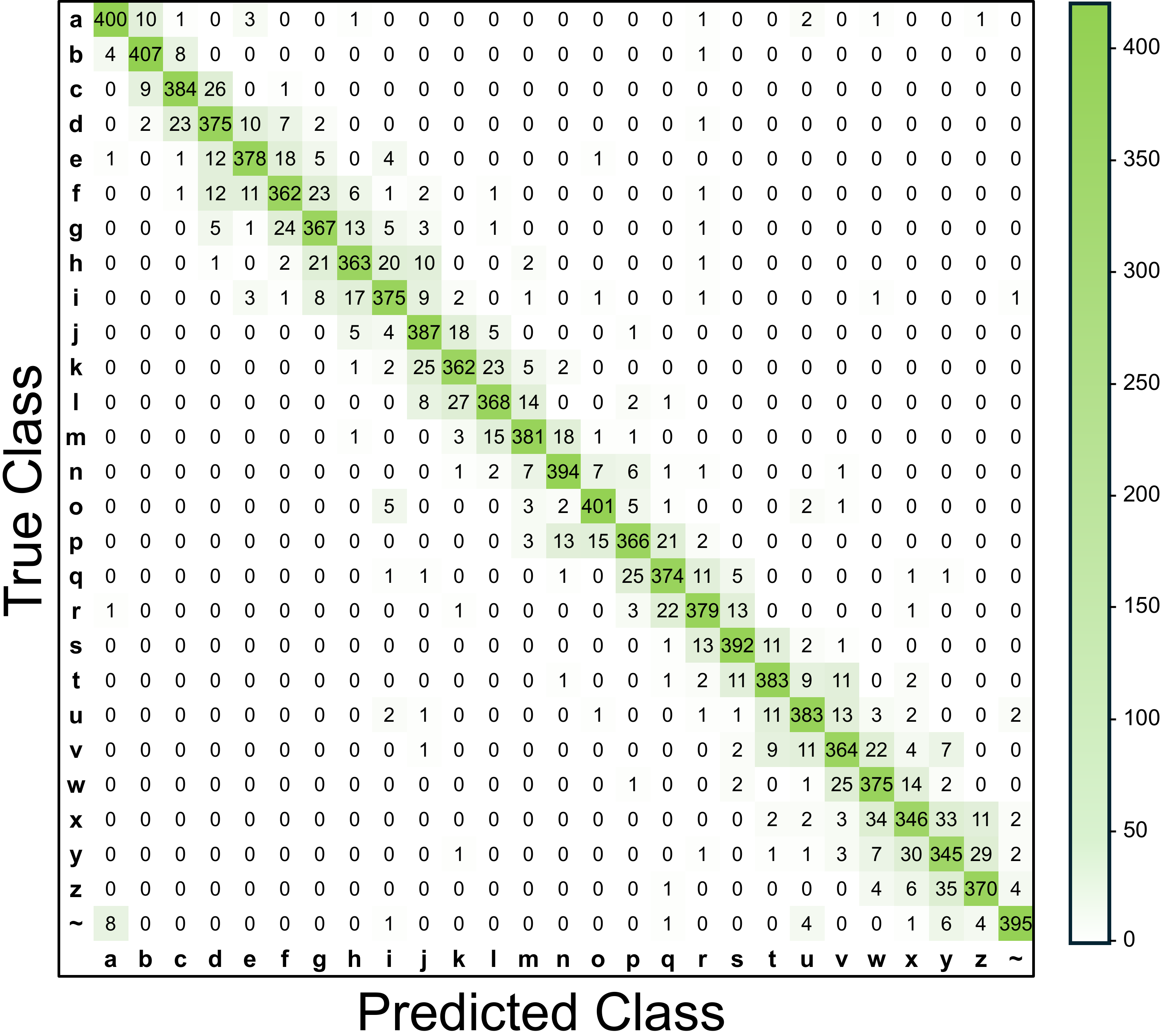}
    \caption{\black{Confusion matrix of the proposed EEdGeNet model evaluated across all 27 classes using 85 extracted features on test data from 15 participants. The do nothing class is denoted by “\textasciitilde”. Most imagined characters are occasionally confused with their neighboring letters, with imagined handwritten character ‘b’ showing the highest classification accuracy and character ‘y’ exhibiting the highest misclassification rate across participants.}}
    \label{fig:confusion_matrix}
\end{figure}

\textcolor{black}{This study establishes the feasibility of real-time decoding of imagined handwriting from non-invasive EEG recordings through a lightweight pipeline designed for efficient deployment on edge devices. As shown in Figure~\ref{fig:confusion_matrix}, most imagined characters are correctly classified, with strong diagonal dominance in the confusion matrix. Certain letters, such as ‘b’, achieve consistently high classification accuracy, whereas others, such as ‘y’, remain challenging and exhibit higher rates of misclassification. Occasional confusions with neighboring letters suggest overlapping neural representations, motivating future work on class-specific feature refinement and adaptive modeling strategies. While within-subject performance is strong, cross-subject generalization remains limited. Leave-one-subject-out (LOSO) experiments yielded an average accuracy of only $3.55 \pm 1.15\%$, with precision and recall showing similarly low values, reflecting substantial inter-subject variability. Addressing this limitation will be a key direction, with emphasis on transfer learning, adaptive calibration, and more flexible architectures to improve subject-independent decoding.}  

\textcolor{black}{Experimental results and comparative evaluations further highlight the trade-offs between accuracy, latency, and energy efficiency. The proposed pipeline for imagined handwriting recognition from EEG—simplified ASR with 85 handcrafted features and the EEdGeNet model—achieved the highest accuracy ($89.83 \pm 0.19\%$) while maintaining favorable inference time and energy balance, underscoring its suitability for real-time edge deployment. Although challenges remain in cross-subject generalization and decoding of inherently ambiguous characters, the proposed framework sets a benchmark for accuracy, efficiency, and deployability, paving the way toward practical neural communication systems.}  

\section*{Conclusion}
This work addresses two fundamental challenges in non-invasive neural decoding: the low signal quality of EEG and the difficulty of deploying real-time models on resource-constrained edge devices. We presented the first real-time demonstration of high-accuracy decoding of imagined handwriting from non-invasive EEG signals using a portable edge device. \black{Our proposed framework combines efficient feature engineering with a hybrid TCN-MLP architecture to enable low-latency inference on resource-constrained hardware, achieving $89.83\% \pm 0.19\%$ classification accuracy with a per-character inference latency of only 914.18 ms on the NVIDIA Jetson TX2.} Through comprehensive evaluation across multiple participants, we demonstrated the robustness and generalizability of the system, highlighting its potential for scalable deployment in practical BCI applications. \black{Importantly, our analysis shows that a carefully selected subset of just ten features—Hjorth mobility, root mean square, Hjorth complexity, delta power, graphical\_A ellipse area, mean dominant frequency, graphical\_D3 ellipse area, graphical\_D1 ellipse area, graphical\_D2 ellipse area, and graphical\_D4 ellipse area—can deliver nearly equivalent accuracy to the full 85-feature set while reducing inference latency by $4.51\times$.} This trade-off between accuracy and efficiency underscores the viability of our system for real-time, on-device neural decoding in assistive communication settings.
However, our results stem from an optimized preprocessing and feature extraction pipeline combined with a lightweight neural model (EEdGeNet), reducing the computational overhead typical of motor imagery classifiers—though outcomes may vary across different applications. Future directions will include expanding the dataset to encompass more participants and diverse recording conditions, which will be critical for enhancing robustness and generalizability. Additional efforts will focus on incorporating hybrid feature sets and adaptive deep learning strategies to improve cross-subject decoding performance.
\section*{Data availability}
The datasets and code used in this study are available from the corresponding author upon reasonable request.


\bibliography{bibiliography}

\begin{thebibliography}{10}
\urlstyle{rm}
\expandafter\ifx\csname url\endcsname\relax
  \def\url#1{\texttt{#1}}\fi
\expandafter\ifx\csname urlprefix\endcsname\relax\def\urlprefix{URL }\fi
\expandafter\ifx\csname doiprefix\endcsname\relax\def\doiprefix{DOI: }\fi
\providecommand{\bibinfo}[2]{#2}
\providecommand{\eprint}[2][]{\url{#2}}

\bibitem{zhang2024brain}
\bibinfo{author}{Zhang, H.} \emph{et~al.}
\newblock \bibinfo{journal}{\bibinfo{title}{Brain--computer interfaces: the innovative key to unlocking neurological conditions}}.
\newblock {\emph{\JournalTitle{International Journal of Surgery}}} \textbf{\bibinfo{volume}{110}}, \bibinfo{pages}{5745--5762} (\bibinfo{year}{2024}).

\bibitem{liu2025recent}
\bibinfo{author}{Liu, X.-Y.} \emph{et~al.}
\newblock \bibinfo{journal}{\bibinfo{title}{Recent applications of eeg-based brain-computer-interface in the medical field}}.
\newblock {\emph{\JournalTitle{Military Medical Research}}} \textbf{\bibinfo{volume}{12}}, \bibinfo{pages}{14} (\bibinfo{year}{2025}).

\bibitem{sun2021hybrid}
\bibinfo{author}{Sun, J.} \emph{et~al.}
\newblock \bibinfo{journal}{\bibinfo{title}{A hybrid deep neural network for classification of schizophrenia using eeg data}}.
\newblock {\emph{\JournalTitle{Scientific Reports}}} \textbf{\bibinfo{volume}{11}}, \bibinfo{pages}{4706} (\bibinfo{year}{2021}).

\bibitem{ahmadlou2012fractality}
\bibinfo{author}{Ahmadlou, M.}, \bibinfo{author}{Adeli, H.} \& \bibinfo{author}{Adeli, A.}
\newblock \bibinfo{journal}{\bibinfo{title}{Fractality analysis of frontal brain in major depressive disorder}}.
\newblock {\emph{\JournalTitle{International Journal of Psychophysiology}}} \textbf{\bibinfo{volume}{85}}, \bibinfo{pages}{206--211} (\bibinfo{year}{2012}).

\bibitem{janapa2022mlperf}
\bibinfo{author}{Janapa~Reddi, V.} \emph{et~al.}
\newblock \bibinfo{journal}{\bibinfo{title}{Mlperf mobile inference benchmark: An industry-standard open-source machine learning benchmark for on-device ai}}.
\newblock {\emph{\JournalTitle{Proceedings of Machine Learning and Systems}}} \textbf{\bibinfo{volume}{4}}, \bibinfo{pages}{352--369} (\bibinfo{year}{2022}).

\bibitem{hadidi2020toward}
\bibinfo{author}{Hadidi, R.}, \bibinfo{author}{Cao, J.}, \bibinfo{author}{Ryoo, M.~S.} \& \bibinfo{author}{Kim, H.}
\newblock \bibinfo{journal}{\bibinfo{title}{Toward collaborative inferencing of deep neural networks on internet-of-things devices}}.
\newblock {\emph{\JournalTitle{IEEE Internet of Things Journal}}} \textbf{\bibinfo{volume}{7}}, \bibinfo{pages}{4950--4960} (\bibinfo{year}{2020}).

\bibitem{safari2024classification}
\bibinfo{author}{Safari, M.}, \bibinfo{author}{Shalbaf, R.}, \bibinfo{author}{Bagherzadeh, S.} \& \bibinfo{author}{Shalbaf, A.}
\newblock \bibinfo{journal}{\bibinfo{title}{Classification of mental workload using brain connectivity and machine learning on electroencephalogram data}}.
\newblock {\emph{\JournalTitle{Scientific Reports}}} \textbf{\bibinfo{volume}{14}}, \bibinfo{pages}{9153} (\bibinfo{year}{2024}).

\bibitem{petrosyan2022speech}
\bibinfo{author}{Petrosyan, A.} \emph{et~al.}
\newblock \bibinfo{journal}{\bibinfo{title}{Speech decoding from a small set of spatially segregated minimally invasive intracranial eeg electrodes with a compact and interpretable neural network}}.
\newblock {\emph{\JournalTitle{Journal of Neural Engineering}}} \textbf{\bibinfo{volume}{19}}, \bibinfo{pages}{066016} (\bibinfo{year}{2022}).

\bibitem{liu2022intracranial}
\bibinfo{author}{Liu, D.} \emph{et~al.}
\newblock \bibinfo{journal}{\bibinfo{title}{Intracranial brain-computer interface spelling using localized visual motion response}}.
\newblock {\emph{\JournalTitle{Neuroimage}}} \textbf{\bibinfo{volume}{258}}, \bibinfo{pages}{119363} (\bibinfo{year}{2022}).

\bibitem{metzger2023high}
\bibinfo{author}{Metzger, S.~L.} \emph{et~al.}
\newblock \bibinfo{journal}{\bibinfo{title}{A high-performance neuroprosthesis for speech decoding and avatar control}}.
\newblock {\emph{\JournalTitle{Nature}}} \textbf{\bibinfo{volume}{620}}, \bibinfo{pages}{1037--1046} (\bibinfo{year}{2023}).

\bibitem{willett2021high}
\bibinfo{author}{Willett, F.~R.}, \bibinfo{author}{Avansino, D.~T.}, \bibinfo{author}{Hochberg, L.~R.}, \bibinfo{author}{Henderson, J.~M.} \& \bibinfo{author}{Shenoy, K.~V.}
\newblock \bibinfo{journal}{\bibinfo{title}{High-performance brain-to-text communication via handwriting}}.
\newblock {\emph{\JournalTitle{Nature}}} \textbf{\bibinfo{volume}{593}}, \bibinfo{pages}{249--254} (\bibinfo{year}{2021}).

\bibitem{mishra2024signeeg}
\bibinfo{author}{Mishra, A.~R.} \emph{et~al.}
\newblock \bibinfo{journal}{\bibinfo{title}{Signeeg v1. 0: Multimodal dataset with electroencephalography and hand-written signature for biometric systems}}.
\newblock {\emph{\JournalTitle{Scientific Data}}} \textbf{\bibinfo{volume}{11}}, \bibinfo{pages}{718} (\bibinfo{year}{2024}).

\bibitem{zhao2021classification}
\bibinfo{author}{Zhao, X.} \emph{et~al.}
\newblock \bibinfo{journal}{\bibinfo{title}{Classification of sleep apnea based on eeg sub-band signal characteristics}}.
\newblock {\emph{\JournalTitle{Scientific Reports}}} \textbf{\bibinfo{volume}{11}}, \bibinfo{pages}{5824} (\bibinfo{year}{2021}).

\bibitem{kim2024electroencephalography}
\bibinfo{author}{Kim, S.-K.}, \bibinfo{author}{Kim, H.}, \bibinfo{author}{Kim, S.~H.}, \bibinfo{author}{Kim, J.~B.} \& \bibinfo{author}{Kim, L.}
\newblock \bibinfo{journal}{\bibinfo{title}{Electroencephalography-based classification of alzheimer’s disease spectrum during computer-based cognitive testing}}.
\newblock {\emph{\JournalTitle{Scientific Reports}}} \textbf{\bibinfo{volume}{14}}, \bibinfo{pages}{5252} (\bibinfo{year}{2024}).

\bibitem{amrani2024deep}
\bibinfo{author}{Amrani, H.}, \bibinfo{author}{Micucci, D.} \& \bibinfo{author}{Napoletano, P.}
\newblock \bibinfo{journal}{\bibinfo{title}{Deep representation learning for open vocabulary electroencephalography-to-text decoding}}.
\newblock {\emph{\JournalTitle{IEEE Journal of Biomedical and Health Informatics}}}  (\bibinfo{year}{2024}).

\bibitem{islam2022explainable}
\bibinfo{author}{Islam, M.~S.}, \bibinfo{author}{Hussain, I.}, \bibinfo{author}{Rahman, M.~M.}, \bibinfo{author}{Park, S.~J.} \& \bibinfo{author}{Hossain, M.~A.}
\newblock \bibinfo{title}{Explainable artificial intelligence model for stroke prediction using eeg signal} (\bibinfo{year}{2022}).

\bibitem{ullah2021imagined}
\bibinfo{author}{Ullah, S.} \& \bibinfo{author}{Halim, Z.}
\newblock \bibinfo{journal}{\bibinfo{title}{Imagined character recognition through eeg signals using deep convolutional neural network}}.
\newblock {\emph{\JournalTitle{Medical \& Biological Engineering \& Computing}}} \textbf{\bibinfo{volume}{59}}, \bibinfo{pages}{1167--1183} (\bibinfo{year}{2021}).

\bibitem{alonso2023eeg}
\bibinfo{author}{Alonso-V{\'a}zquez, D.}, \bibinfo{author}{Mendoza-Montoya, O.}, \bibinfo{author}{Caraza, R.}, \bibinfo{author}{Martinez, H.~R.} \& \bibinfo{author}{Antelis, J.~M.}
\newblock \bibinfo{journal}{\bibinfo{title}{Eeg-based classification of spoken words using machine learning approaches}}.
\newblock {\emph{\JournalTitle{Computation}}} \textbf{\bibinfo{volume}{11}}, \bibinfo{pages}{225} (\bibinfo{year}{2023}).

\bibitem{abdulghani2023imagined}
\bibinfo{author}{Abdulghani, M.~M.}, \bibinfo{author}{Walters, W.~L.} \& \bibinfo{author}{Abed, K.~H.}
\newblock \bibinfo{journal}{\bibinfo{title}{Imagined speech classification using eeg and deep learning}}.
\newblock {\emph{\JournalTitle{Bioengineering}}} \textbf{\bibinfo{volume}{10}}, \bibinfo{pages}{649} (\bibinfo{year}{2023}).

\bibitem{panachakel2021decoding}
\bibinfo{author}{Panachakel, J.~T.} \& \bibinfo{author}{Ganesan, R.~A.}
\newblock \bibinfo{journal}{\bibinfo{title}{Decoding imagined speech from eeg using transfer learning}}.
\newblock {\emph{\JournalTitle{IEEE Access}}} \textbf{\bibinfo{volume}{9}}, \bibinfo{pages}{135371--135383} (\bibinfo{year}{2021}).

\bibitem{altaheri2022physics}
\bibinfo{author}{Altaheri, H.}, \bibinfo{author}{Muhammad, G.} \& \bibinfo{author}{Alsulaiman, M.}
\newblock \bibinfo{journal}{\bibinfo{title}{Physics-informed attention temporal convolutional network for eeg-based motor imagery classification}}.
\newblock {\emph{\JournalTitle{IEEE transactions on industrial informatics}}} \textbf{\bibinfo{volume}{19}}, \bibinfo{pages}{2249--2258} (\bibinfo{year}{2022}).

\bibitem{xie2022transformer}
\bibinfo{author}{Xie, J.} \emph{et~al.}
\newblock \bibinfo{journal}{\bibinfo{title}{A transformer-based approach combining deep learning network and spatial-temporal information for raw eeg classification}}.
\newblock {\emph{\JournalTitle{IEEE Transactions on Neural Systems and Rehabilitation Engineering}}} \textbf{\bibinfo{volume}{30}}, \bibinfo{pages}{2126--2136} (\bibinfo{year}{2022}).

\bibitem{khademi2022transfer}
\bibinfo{author}{Khademi, Z.}, \bibinfo{author}{Ebrahimi, F.} \& \bibinfo{author}{Kordy, H.~M.}
\newblock \bibinfo{journal}{\bibinfo{title}{A transfer learning-based cnn and lstm hybrid deep learning model to classify motor imagery eeg signals}}.
\newblock {\emph{\JournalTitle{Computers in biology and medicine}}} \textbf{\bibinfo{volume}{143}}, \bibinfo{pages}{105288} (\bibinfo{year}{2022}).

\bibitem{cho2021neurograsp}
\bibinfo{author}{Cho, J.-H.}, \bibinfo{author}{Jeong, J.-H.} \& \bibinfo{author}{Lee, S.-W.}
\newblock \bibinfo{journal}{\bibinfo{title}{Neurograsp: Real-time eeg classification of high-level motor imagery tasks using a dual-stage deep learning framework}}.
\newblock {\emph{\JournalTitle{IEEE Transactions on Cybernetics}}} \textbf{\bibinfo{volume}{52}}, \bibinfo{pages}{13279--13292} (\bibinfo{year}{2021}).

\bibitem{nakanishi2017enhancing}
\bibinfo{author}{Nakanishi, M.} \emph{et~al.}
\newblock \bibinfo{journal}{\bibinfo{title}{Enhancing detection of ssveps for a high-speed brain speller using task-related component analysis}}.
\newblock {\emph{\JournalTitle{IEEE Transactions on Biomedical Engineering}}} \textbf{\bibinfo{volume}{65}}, \bibinfo{pages}{104--112} (\bibinfo{year}{2017}).

\bibitem{farwell1988talking}
\bibinfo{author}{Farwell, L.~A.} \& \bibinfo{author}{Donchin, E.}
\newblock \bibinfo{journal}{\bibinfo{title}{Talking off the top of your head: toward a mental prosthesis utilizing event-related brain potentials}}.
\newblock {\emph{\JournalTitle{Electroencephalography and clinical Neurophysiology}}} \textbf{\bibinfo{volume}{70}}, \bibinfo{pages}{510--523} (\bibinfo{year}{1988}).

\bibitem{huotari2019sampling}
\bibinfo{author}{Huotari, N.} \emph{et~al.}
\newblock \bibinfo{journal}{\bibinfo{title}{Sampling rate effects on resting state fmri metrics}}.
\newblock {\emph{\JournalTitle{Frontiers in neuroscience}}} \textbf{\bibinfo{volume}{13}}, \bibinfo{pages}{279} (\bibinfo{year}{2019}).

\bibitem{beckmann2005investigations}
\bibinfo{author}{Beckmann, C.~F.}, \bibinfo{author}{DeLuca, M.}, \bibinfo{author}{Devlin, J.~T.} \& \bibinfo{author}{Smith, S.~M.}
\newblock \bibinfo{journal}{\bibinfo{title}{Investigations into resting-state connectivity using independent component analysis}}.
\newblock {\emph{\JournalTitle{Philosophical Transactions of the Royal Society B: Biological Sciences}}} \textbf{\bibinfo{volume}{360}}, \bibinfo{pages}{1001--1013} (\bibinfo{year}{2005}).

\bibitem{chang2019evaluation}
\bibinfo{author}{Chang, C.-Y.}, \bibinfo{author}{Hsu, S.-H.}, \bibinfo{author}{Pion-Tonachini, L.} \& \bibinfo{author}{Jung, T.-P.}
\newblock \bibinfo{journal}{\bibinfo{title}{Evaluation of artifact subspace reconstruction for automatic artifact components removal in multi-channel eeg recordings}}.
\newblock {\emph{\JournalTitle{IEEE transactions on biomedical engineering}}} \textbf{\bibinfo{volume}{67}}, \bibinfo{pages}{1114--1121} (\bibinfo{year}{2019}).

\bibitem{blum2019riemannian}
\bibinfo{author}{Blum, S.}, \bibinfo{author}{Jacobsen, N.~S.}, \bibinfo{author}{Bleichner, M.~G.} \& \bibinfo{author}{Debener, S.}
\newblock \bibinfo{journal}{\bibinfo{title}{A riemannian modification of artifact subspace reconstruction for eeg artifact handling}}.
\newblock {\emph{\JournalTitle{Frontiers in human neuroscience}}} \textbf{\bibinfo{volume}{13}}, \bibinfo{pages}{141} (\bibinfo{year}{2019}).

\bibitem{Yeh2025Corticomuscular}
\bibinfo{author}{Yeh, C.} \& \bibinfo{author}{colleagues}.
\newblock \bibinfo{journal}{\bibinfo{title}{Corticomuscular coherence study}}.
\newblock {\emph{\JournalTitle{Scientific Reports}}}  (\bibinfo{year}{2025}).
\newblock \bibinfo{note}{Band-pass filtering EEG 1–50 Hz with zero-phase FIR filter}.

\bibitem{Mahadevan2024EEGfMRI}
\bibinfo{author}{Mahadevan, R.} \& \bibinfo{author}{colleagues}.
\newblock \bibinfo{journal}{\bibinfo{title}{Eeg–fmri emotion recognition pipeline}}.
\newblock {\emph{\JournalTitle{Frontiers in Computational Neuroscience}}}  (\bibinfo{year}{2024}).
\newblock \bibinfo{note}{EEG band-pass filtered 1–50 Hz before artifact removal}.

\bibitem{Choi2018EarEEG}
\bibinfo{author}{Choi, J.} \& \bibinfo{author}{colleagues}.
\newblock \bibinfo{journal}{\bibinfo{title}{Ear-eeg brain–computer interface methods}}.
\newblock {\emph{\JournalTitle{Sensors}}} \textbf{\bibinfo{volume}{18}} (\bibinfo{year}{2018}).
\newblock \bibinfo{note}{4th-order Butterworth band-pass 1–50 Hz}.

\bibitem{mullen2015real}
\bibinfo{author}{Mullen, T.~R.} \emph{et~al.}
\newblock \bibinfo{journal}{\bibinfo{title}{Real-time neuroimaging and cognitive monitoring using wearable dry eeg}}.
\newblock {\emph{\JournalTitle{IEEE transactions on biomedical engineering}}} \textbf{\bibinfo{volume}{62}}, \bibinfo{pages}{2553--2567} (\bibinfo{year}{2015}).

\bibitem{akbari2023recognizing}
\bibinfo{author}{Akbari, H.} \emph{et~al.}
\newblock \bibinfo{journal}{\bibinfo{title}{Recognizing seizure using poincar{\'e} plot of eeg signals and graphical features in dwt domain}}.
\newblock {\emph{\JournalTitle{Bratislava Medical Journal/Bratislavske Lekarske Listy}}} \textbf{\bibinfo{volume}{124}}, \bibinfo{pages}{12--24} (\bibinfo{year}{2023}).

\bibitem{hjorth1970eeg}
\bibinfo{author}{Hjorth, B.}
\newblock \bibinfo{journal}{\bibinfo{title}{Eeg analysis based on time domain properties}}.
\newblock {\emph{\JournalTitle{Electroencephalography and clinical neurophysiology}}} \textbf{\bibinfo{volume}{29}}, \bibinfo{pages}{306--310} (\bibinfo{year}{1970}).

\bibitem{bandt2002permutation}
\bibinfo{author}{Bandt, C.} \& \bibinfo{author}{Pompe, B.}
\newblock \bibinfo{journal}{\bibinfo{title}{Permutation entropy: a natural complexity measure for time series}}.
\newblock {\emph{\JournalTitle{Physical review letters}}} \textbf{\bibinfo{volume}{88}}, \bibinfo{pages}{174102} (\bibinfo{year}{2002}).

\bibitem{pernagallo2025random}
\bibinfo{author}{Pernagallo, G.}
\newblock \bibinfo{journal}{\bibinfo{title}{Random walks, hurst exponent, and market efficiency: G. pernagallo}}.
\newblock {\emph{\JournalTitle{Quality \& Quantity}}} \textbf{\bibinfo{volume}{59}}, \bibinfo{pages}{1097--1119} (\bibinfo{year}{2025}).

\bibitem{sadiq2022alcoholic}
\bibinfo{author}{Sadiq, M.~T.}, \bibinfo{author}{Akbari, H.}, \bibinfo{author}{Siuly, S.}, \bibinfo{author}{Li, Y.} \& \bibinfo{author}{Wen, P.}
\newblock \bibinfo{journal}{\bibinfo{title}{Alcoholic eeg signals recognition based on phase space dynamic and geometrical features}}.
\newblock {\emph{\JournalTitle{Chaos, solitons \& fractals}}} \textbf{\bibinfo{volume}{158}}, \bibinfo{pages}{112036} (\bibinfo{year}{2022}).

\bibitem{akbari2021depression}
\bibinfo{author}{Akbari, H.} \emph{et~al.}
\newblock \bibinfo{journal}{\bibinfo{title}{Depression recognition based on the reconstruction of phase space of eeg signals and geometrical features}}.
\newblock {\emph{\JournalTitle{Applied Acoustics}}} \textbf{\bibinfo{volume}{179}}, \bibinfo{pages}{108078} (\bibinfo{year}{2021}).

\bibitem{bai2018empirical}
\bibinfo{author}{Bai, S.}, \bibinfo{author}{Kolter, J.~Z.} \& \bibinfo{author}{Koltun, V.}
\newblock \bibinfo{journal}{\bibinfo{title}{An empirical evaluation of generic convolutional and recurrent networks for sequence modeling}}.
\newblock {\emph{\JournalTitle{arXiv preprint arXiv:1803.01271}}}  (\bibinfo{year}{2018}).

\bibitem{sadiq2022motor}
\bibinfo{author}{Sadiq, M.~T.} \emph{et~al.}
\newblock \bibinfo{journal}{\bibinfo{title}{Motor imagery bci classification based on multivariate variational mode decomposition}}.
\newblock {\emph{\JournalTitle{IEEE Transactions on Emerging Topics in Computational Intelligence}}} \textbf{\bibinfo{volume}{6}}, \bibinfo{pages}{1177--1189} (\bibinfo{year}{2022}).

\bibitem{sadiq2019motor}
\bibinfo{author}{Sadiq, M.~T.} \emph{et~al.}
\newblock \bibinfo{journal}{\bibinfo{title}{Motor imagery eeg signals decoding by multivariate empirical wavelet transform-based framework for robust brain--computer interfaces}}.
\newblock {\emph{\JournalTitle{IEEE access}}} \textbf{\bibinfo{volume}{7}}, \bibinfo{pages}{171431--171451} (\bibinfo{year}{2019}).

\bibitem{sadiq2024fast}
\bibinfo{author}{Sadiq, M.~T.}, \bibinfo{author}{Yousaf, A.}, \bibinfo{author}{Siuly, S.} \& \bibinfo{author}{Almogren, A.}
\newblock \bibinfo{journal}{\bibinfo{title}{Fast fractional fourier transform-aided novel graphical approach for eeg alcoholism detection}}.
\newblock {\emph{\JournalTitle{Bioengineering}}} \textbf{\bibinfo{volume}{11}}, \bibinfo{pages}{464} (\bibinfo{year}{2024}).

\bibitem{sadiq2022exploiting}
\bibinfo{author}{Sadiq, M.~T.} \emph{et~al.}
\newblock \bibinfo{journal}{\bibinfo{title}{Exploiting pretrained cnn models for the development of an eeg-based robust bci framework}}.
\newblock {\emph{\JournalTitle{Computers in Biology and Medicine}}} \textbf{\bibinfo{volume}{143}}, \bibinfo{pages}{105242} (\bibinfo{year}{2022}).

\bibitem{schirrmeister2017deep}
\bibinfo{author}{Schirrmeister, R.~T.} \emph{et~al.}
\newblock \bibinfo{journal}{\bibinfo{title}{Deep learning with convolutional neural networks for eeg decoding and visualization}}.
\newblock {\emph{\JournalTitle{Human brain mapping}}} \textbf{\bibinfo{volume}{38}}, \bibinfo{pages}{5391--5420} (\bibinfo{year}{2017}).

\bibitem{lawhern2018eegnet}
\bibinfo{author}{Lawhern, V.~J.} \emph{et~al.}
\newblock \bibinfo{journal}{\bibinfo{title}{Eegnet: a compact convolutional neural network for eeg-based brain--computer interfaces}}.
\newblock {\emph{\JournalTitle{Journal of neural engineering}}} \textbf{\bibinfo{volume}{15}}, \bibinfo{pages}{056013} (\bibinfo{year}{2018}).

\bibitem{ingolfsson2020eeg}
\bibinfo{author}{Ingolfsson, T.~M.} \emph{et~al.}
\newblock \bibinfo{title}{Eeg-tcnet: An accurate temporal convolutional network for embedded motor-imagery brain--machine interfaces}.
\newblock In \emph{\bibinfo{booktitle}{2020 IEEE International Conference on Systems, Man, and Cybernetics (SMC)}}, \bibinfo{pages}{2958--2965} (\bibinfo{organization}{IEEE}, \bibinfo{year}{2020}).

\bibitem{musallam2021electroencephalography}
\bibinfo{author}{Musallam, Y.~K.} \emph{et~al.}
\newblock \bibinfo{journal}{\bibinfo{title}{Electroencephalography-based motor imagery classification using temporal convolutional network fusion}}.
\newblock {\emph{\JournalTitle{Biomedical Signal Processing and Control}}} \textbf{\bibinfo{volume}{69}}, \bibinfo{pages}{102826} (\bibinfo{year}{2021}).

\bibitem{altuwaijri2022multi}
\bibinfo{author}{Altuwaijri, G.~A.}, \bibinfo{author}{Muhammad, G.}, \bibinfo{author}{Altaheri, H.} \& \bibinfo{author}{Alsulaiman, M.}
\newblock \bibinfo{journal}{\bibinfo{title}{A multi-branch convolutional neural network with squeeze-and-excitation attention blocks for eeg-based motor imagery signals classification}}.
\newblock {\emph{\JournalTitle{Diagnostics}}} \textbf{\bibinfo{volume}{12}}, \bibinfo{pages}{995} (\bibinfo{year}{2022}).

\bibitem{yu2021new}
\bibinfo{author}{Yu, X.}, \bibinfo{author}{Aziz, M.~Z.}, \bibinfo{author}{Sadiq, M.~T.}, \bibinfo{author}{Fan, Z.} \& \bibinfo{author}{Xiao, G.}
\newblock \bibinfo{journal}{\bibinfo{title}{A new framework for automatic detection of motor and mental imagery eeg signals for robust bci systems}}.
\newblock {\emph{\JournalTitle{IEEE Transactions on Instrumentation and Measurement}}} \textbf{\bibinfo{volume}{70}}, \bibinfo{pages}{1--12} (\bibinfo{year}{2021}).

\bibitem{skomrock2018characterization}
\bibinfo{author}{Skomrock, N.~D.} \emph{et~al.}
\newblock \bibinfo{journal}{\bibinfo{title}{A characterization of brain-computer interface performance trade-offs using support vector machines and deep neural networks to decode movement intent}}.
\newblock {\emph{\JournalTitle{Frontiers in neuroscience}}} \textbf{\bibinfo{volume}{12}}, \bibinfo{pages}{763} (\bibinfo{year}{2018}).

\end{thebibliography}





\section*{Author contributions statement}

B.C. and O.S. conceptualized the idea. O.S., R.S., and D.V. were responsible for participant recruitment and data acquisition. O.S. and D.V. led the feature extraction and selection process. R.S. and O.S. performed machine learning model analysis and performance evaluation. D.V. and A.P. contributed to channel selection and spatial electrode configuration. S.J. conducted dataset-level analyses. P.L. provided input on experimental design. A.Ka. and A.Kh. contributed to the technical discussions. O.S. and R.S. wrote the main manuscript. B.C. supervised the project and provided overall research direction and guidance. All authors reviewed and approved the final manuscript. 

\section*{Funding}
This research received no specific grant from any funding agency in the public, commercial, or not-for-profit sectors.

\section*{Additional information}
Correspondence and requests for materials should be addressed to O.S.






\end{document}